\documentclass{article}
\usepackage[margin=1in]{geometry}
\usepackage{booktabs}
\usepackage{longtable}
\usepackage{array}
\usepackage{multirow}
\usepackage{xcolor}
\usepackage{colortbl}
\usepackage{float}
\usepackage{amsmath}
\usepackage{amssymb}
\usepackage{caption}
\usepackage{subcaption}
\usepackage{dsfont}
\usepackage{graphicx}
\usepackage[numbers]{natbib}
\usepackage[hyphens]{url}
\usepackage{hyperref}

\newcolumntype{L}{>{\raggedright\arraybackslash}X}
\newcolumntype{R}{>{\raggedleft\arraybackslash}X}
\newcolumntype{C}{>{\centering\arraybackslash}X}

\title{A model predictive control framework with customer-priority tiers for virtual power plant resilience during extreme weather: A UK heatwave case study}

\author{Edward Moroshko, Weizhe Qin, Desen Kirli,\\ Mohammed Qais, Sotirios Tsaftaris, Aristides Kiprakis}
\date{}

\begin{document}

\maketitle


\begin{abstract}
Due to changes in frequency and intensity of extreme weather events, such as heatwaves and storms, power systems around the globe are having to deal with increased imbalance between demand and supply and additional risk of loss of supply, calling for advanced control strategies that strengthen system resilience. This paper develops a Model Predictive Control (MPC) framework for coordination of Virtual Power Plants (VPPs) that manages photovoltaic (PV) systems, batteries, and loads before, during, and after extreme weather events. A multi-objective mixed-integer quadratically constrained program is solved to enforce customer-priority tiers, serving critical loads first, while minimizing operating cost and PV curtailment under network and device constraints. Simulations on the IEEE 33-bus distribution network with real UK heatwave data show that, under realistic forecast errors and modeling uncertainties, MPC improves resilience by $11\text{--}20\%$ relative to traditional full-horizon optimization. These results indicate the practical viability of receding-horizon coordination for resilient, low-carbon VPP operation during extreme weather.
\end{abstract}

\section{Introduction}
The increasing frequency and intensity of extreme weather events due to climate change pose unprecedented challenges to electrical power systems worldwide. 
Traditional centralized power generation and distribution systems are particularly vulnerable to severe weather phenomena such as hurricanes, heatwaves, and ice storms, which can cause widespread power outages and significant economic losses \cite{7036086, 8974395, PANTELI2015259}. These events often trigger cascading failures that leave millions of customers without power for extended periods.
For example, 
Winter Storm Uri in February 2021 caused cascading failures across Texas, affecting $4.5$ million homes and businesses, resulting in $\$195-295$ billion in damages \cite{FERC2021Uri}.
Storms Dudley, Eunice, and Franklin in February 2022 caused widespread power outages across the UK and Northern Europe, leaving over $1$ million homes without electricity and causing significant infrastructure damage and economic disruption \cite{MetOffice2022_DudleyEuniceFranklin}.

Virtual Power Plants (VPPs) have emerged as a promising solution to enhance grid resilience during such extreme weather events. VPPs aggregate distributed energy resources (DERs) including renewable generation, energy storage systems, and flexible loads under coordinated control, effectively operating as a single flexible plant \cite{10415859, 8881433}. During normal conditions, VPPs support grid stability, but during emergencies, they can transition to islanded operation, creating self-sustaining systems that can operate independently when the main grid becomes unavailable.
While real-world VPP deployments exist, such as Europe's Next Kraftwerke \cite{NextKraftwerke2022} and Tesla's South Australian VPP \cite{cefc_savpp_2023}, such implementations remain scarce and primarily focus on economic optimization under normal grid conditions.
As extreme weather events become increasingly frequent, there is a critical need to develop VPPs specifically for resilience applications through new theoretical frameworks and algorithms.
Existing deployments, while promising, require advanced coordination strategies that can effectively respond to emergency conditions and enhance grid resilience during outages.

Coordinating DERs within VPPs presents complex optimization challenges, particularly during emergency conditions when rapid decision-making is crucial. The core challenge lies in solving a multi-objective dispatch optimization problem in real-time while managing highly uncertain forecasts for renewable generation and load demand \cite{9682593}. In addition, during extended outages, limited local generation may force load shedding, so it is essential to prioritize higher-importance customers (critical and high-priority loads) to maximize the utility of available power. At the same time, when grid power is available (before or after the event), cost-effective operation remains important – for instance, leveraging dynamic pricing to decide how much power to import from the main grid versus using local generation or storage \cite{HERDING2024100180}. Another objective is to maximize the utilization of onsite renewables, such as photovoltaic (PV) systems, both to sustain the system and to avoid curtailing clean energy. These multifaceted goals – resilience, economic cost, and renewable utilization – must be balanced within the VPP's energy management strategy \cite{9061780, NOSRATABADI2017341}.

Existing VPP coordination strategies typically rely on full-horizon optimization methods that assume perfect forecasts over extended periods \cite{Souto2024MPC, harsh2024stochastic, QIN2025111329}. These approaches become computationally intractable for large-scale systems, and they also cannot adapt to dynamic conditions typical during extreme weather events \cite{ROALD2023108725}. Their perceived optimality becomes illusory when confronted with real-world forecast deviations. In scenarios where forecast errors are commonplace, static pre-calculated plans quickly become suboptimal or even infeasible \cite{ZAMANI2016324}.

To address these limitations, we propose a Model Predictive Control (MPC) framework for VPP coordination before, during, and after extreme weather events.
MPC employs a receding-horizon approach that continuously re-optimizes based on updated forecasts and measurements, providing inherent robustness to uncertainties \cite{Schwenzer2021ReviewMPC, HU2021110422}.
This approach is particularly valuable during extreme weather events when forecast accuracy degrades rapidly and system conditions change unpredictably.
Our MPC framework optimizes a multi-objective function that balances resilience with economic efficiency and renewable energy usage, formulated as a Mixed-Integer Quadratically Constrained Programming (MIQCP) problem.

Our proposed approach incorporates several technical innovations to ensure accuracy and realism. First, to improve the accuracy of the power flow model within the MPC loop, we introduce an enhanced delta optimization for linearized distribution flow (LinDistFlow) \cite{9682997}. This approach improves accuracy by optimizing voltage and power flow deltas around the most recent operating point, making it particularly effective for the iterative nature of MPC. Second, we use a time-varying battery efficiency model to capture realistic temperature dependencies \cite{FORGEZ20102961}.

To validate the effectiveness of the proposed approach, we conduct simulations on the IEEE 33-bus distribution network. The validation includes an extreme weather scenario with grid outage, where the main grid is unavailable and the VPP must rely on renewable generation and batteries. We perform a comprehensive comparative analysis between our proposed MPC framework and traditional full-horizon optimization under realistic conditions, including forecasting errors in load demand and renewable generation, as well as modeling uncertainties such as time-varying battery efficiency. The results validate the effectiveness of our proposed approach.

\subsection{Contributions}
Across the literature, MPC is either commercially oriented or invoked only post-event \cite{Souto2024MPC}. No study, to our knowledge, embeds MPC in a Technical VPP (TVPP) throughout the entire extreme-weather timeline. The present work fills this gap by coupling a TVPP with a rolling-horizon controller that continuously operates before, during, and after an event, explicitly balancing resilience, cost, and renewable utilization under network constraints.

To summarize, the main contributions of this work are: 
\begin{itemize}
    \item A multi-objective optimization function with customer-priority tiers that balances resilience with economic efficiency and renewable energy usage (Section \ref{sec:math}).
    \item A novel MPC-based framework specifically designed for VPP resilience before, during, and after extreme weather (Section \ref{sec:mpc_vs_full}).
    \item A comprehensive comparative analysis demonstrating MPC superiority over full-horizon optimization under forecasting errors and modeling uncertainties. These are validated with real UK heatwave data on IEEE 33-bus distribution network (Sections \ref{sec:simulations} and \ref{sec:result_analysis}).
\end{itemize}

The implementation code for this work is publicly available at \url{https://github.com/VPP-WARD/mpc_extreme_weather}.

\subsection{Related Work}
We categorize the related work into three areas based on their approach to VPP coordination and use of MPC. Commercial VPPs (CVPPs) treat the VPP as a market participant that aggregates distributed resources primarily to maximize revenue through electricity trading and ancillary services, with MPC employed for economic optimization under market uncertainties. Technical VPPs (TVPPs) embed detailed distribution network physics into their optimization frameworks, explicitly modeling power flow equations, voltage constraints, and network topology to ensure feasible and reliable operation, with MPC applied to manage DERs in real-time. Finally, VPP approaches for resilience without MPC address extreme weather preparedness through static planning methodologies, typically focusing on optimal placement and sizing of distributed resources to withstand anticipated disruptions, but without employing receding-horizon control strategies.

\subsubsection{MPC in CVPPs}
CVPP applications employ MPC primarily for profit maximization under market uncertainties.
\citet{APPINO2021106738} developed a stochastic MPC that co-optimizes electricity-to-hydrogen conversion and real-time market bids, boosting revenue while respecting electrolyser limits. 
Distributed MPC variants have emerged to address scalability, e.g., \citet{VELASQUEZ2019607} applied dual-decomposition MPC to economic dispatch with high renewable penetration, allowing each DER cluster to optimize locally and exchange only price signals.
Advanced hierarchical designs go further: \citet{9722885} combine multi-timescale Intent Profiles with Dynamic Storage Virtualization so a CVPP can stack revenues across multiple markets. Enhanced MPC schemes also appear: \citet{8912399} embed short-horizon forecast-error correction in a two-stage MPC for an Australian CVPP, cutting supply-demand mismatch in field trials. \citet{MOHAMMADZADEH2022230} proposed an MPC dispatch scheme for a 115 MW solar-tower plant with eight-hour thermal storage, beating day-ahead planning by $8$–$15\%$ under real-time spot-price volatility. These works optimize economics under assumed grid availability, while resilience is outside their scope.

\subsubsection{MPC in Technical VPPs}
A TVPP embeds distribution network physics in its optimization. \citet{9682593} introduced the term and formulated a day-ahead Mixed-Integer Linear Program that limits feeder voltages while scheduling diverse DERs. \citet{LEVIEUX2021119535} applied MPC for joint scheduling of wind farms and reservoir hydro, focusing on maximizing renewable energy penetration rather than maintaining power delivery during disturbances.
The most relevant work addressing extreme weather resilience is by \citet{Souto2024MPC}, who developed an MPC-based framework for TVPP incorporating pre-disaster and post-disaster actions. Their two-stage approach employs Mixed-Integer Quadratic Programming for deterministic pre-disaster planning, followed by MPC for post-disaster corrective actions. The pre-disaster phase uses traditional static optimization over the full horizon, computed once and never revised, with adaptive MPC control beginning only after damage materializes. This framework lacks receding-horizon adaptability during the critical hours before impact, when forecasts and resource availability evolve most rapidly.

\subsubsection{VPP for Resilience Without MPC}
Resilience-focused research without MPC typically addresses the problem as static infrastructure planning. \citet{DEHGHAN20234243} embedded stochastic storm scenarios in a bi-objective siting model that balances annual cost against expected unserved energy. Similar placement studies rely on meta-heuristics: Jellyfish Search \cite{su17031043}, Hunting-Prey Optimization \cite{YUVARAJ20246094}, and a hybrid Crow-Search/Sine-Cosine algorithm \cite{Zadehbagheri2025_resilienceVPP} each produce Pareto-optimal VPP layouts for floods or earthquakes.
A few papers extend the planning perspective to a two-stage response, yet still lack receding-horizon feedback. \citet{MENG2023108918} first harden critical nodes and lines, then apply a static post-event recovery schedule based on a resilience index. \citet{11009860} pre-position drainage crews, repair teams, and mobile storage before typhoon-flood events, dispatching them afterwards according to a deterministic rule set.

More recently, \citet{QIN2025111329} developed a weather-informed optimization framework for EV scheduling under extreme weather in the Scottish islands, employing enhanced Grey Wolf Optimization with stochastic scenario generation and physics-aware reduction to solve a full 24-hour planning problem. While this approach explicitly models storm-induced wind cut-outs and incorporates forecast uncertainty through scenario clustering, the optimization remains full-horizon: decisions are computed once for the entire day based on day-ahead forecasts, without receding-horizon re-optimization as conditions evolve.

While such approaches improve outage recovery capabilities and in some cases incorporate forecast uncertainty, they provide plans computed once for the entire horizon that cannot adapt dynamically as actual conditions deviate from forecasts and system states evolve.

In summary, existing approaches either optimize economic performance under normal conditions or provide static resilience planning, but none addresses the dynamic coordination challenges during evolving extreme weather scenarios through receding-horizon control.

\subsection{Paper Structure}
The remainder of this paper is organized as follows.
Section \ref{sec:system} provides a detailed description of the VPP system components and operational phases.
Section \ref{sec:math} presents the mathematical formulation of the optimization problem, including the constraints and the multi-objective optimization function.
Section \ref{sec:mpc_vs_full} presents detailed descriptions of both the traditional full-horizon optimization and the proposed MPC approach.
Section \ref{uncertain} describes the forecasting uncertainty models and time-varying battery efficiency formulation used to evaluate system performance under realistic conditions. Section \ref{sec:simulations} describes the modeling approach, including input data, parameter selection, and performance metrics.
Section \ref{sec:result_analysis}
presents simulation results and analysis validating the effectiveness of the proposed MPC framework. Finally, Section \ref{sec:conclusions} concludes the paper and discusses future research directions.

\section{System Description}
\label{sec:system}
\begin{figure}[t]
  \centering
  \includegraphics[width=\linewidth]{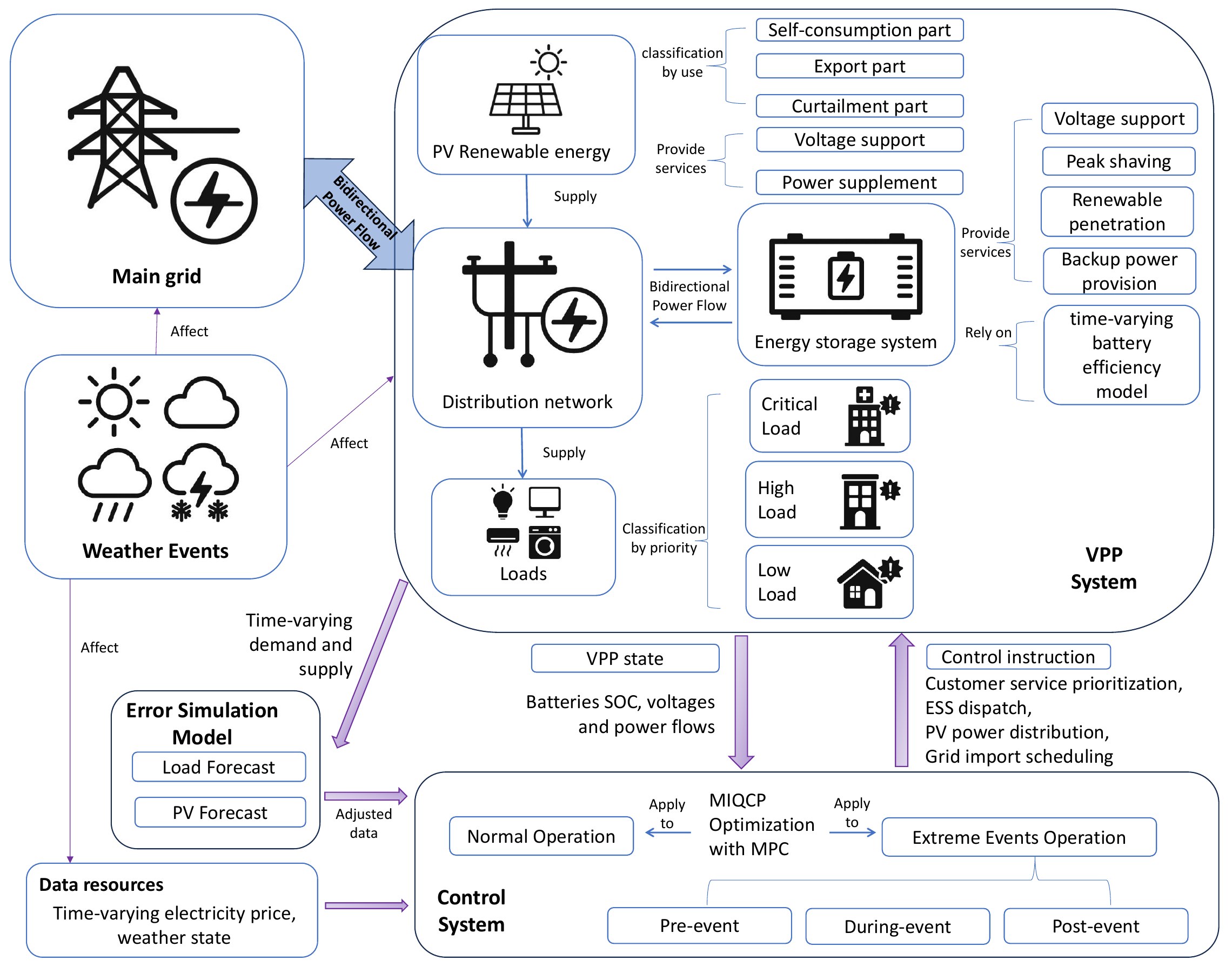}
  \caption{Overall VPP architecture showing physical assets (PV, ESS, loads, distribution network), exogenous inputs (weather and price signals), forecasting/error model, MPC-based control, and optimization core across pre-, during-, and post-event phases.}
  \label{fig:vpp-arch}
\end{figure}

The VPP acts as a centralized coordinator for distributed energy resources connected to a radial distribution network. Figure \ref{fig:vpp-arch} illustrates the complete VPP system architecture. Physical assets include PV systems, battery energy storage, and customer loads (classified as critical, high, and low priority). Weather events affect both the physical distribution network and the generation/demand patterns. An error simulation model produces time-varying demand and PV supply forecasts that incorporate realistic uncertainties. The controller receives these forecasts along with data resources (time-varying electricity prices and weather state information) and current system measurements (bus voltages, branch flows, battery SoC). The controller runs an MPC loop that solves the MIQCP optimization problem, then issues control instructions for customer service prioritization, Energy Storage Systems (ESS) dispatch, PV power distribution, and grid import scheduling across pre-event, during-event, and post-event phases. Our implementation uses the IEEE 33-bus distribution system as a representative testbed, though the framework applies to any radial distribution topology. 

\subsection{System Components}
\label{sec:components}

\textbf{Customer Loads} are distributed across network buses and classified by operational priority to enable systematic load shedding during resource-constrained conditions. Critical priority customers include essential services such as hospitals, emergency facilities, and critical infrastructure that must maintain power supply during extreme events. High priority customers represent important commercial and residential customers whose service significantly impacts community welfare. Low priority customers are standard customers who can be temporarily disconnected to preserve power for higher-priority loads. Each customer exhibits time-varying active power demand and operates at a specified power factor that determines reactive power consumption. The binary served status is part of the decision variables used for customer service prioritization in the optimization model. This implements selective shedding or service guarantees under stressed conditions while keeping the problem tractable.


\textbf{Battery Energy Storage Systems} provide services such as voltage support, peak shaving, renewable penetration improvement, and backup supply to priority loads. As shown in Figure \ref{fig:vpp-arch}, these systems serve dual functions depending on operational phase: peak shaving and renewable integration during normal conditions, and backup power provision during grid outages. 
Battery inverters provide reactive power capability for voltage regulation. The battery SoC evolves according to the discrete-time model introduced in Section~\ref{sec:math}, and a temperature-dependent efficiency model is applied and updated between MPC iterations; details are given in Section~\ref{sec:battery}.



\textbf{PV Systems} are deployed at selected customer locations to provide local renewable generation. As illustrated in Figure \ref{fig:vpp-arch}, each PV system dynamically distributes its time-varying output among three operational modes: self-consumption for direct supply to the customer's local load (reducing net grid demand), export of surplus power injected into the network to serve other customers or charge battery storage systems, and curtailment representing unused generation capacity due to system constraints or excess supply conditions. This three-way split is enforced by the balance and bound constraints in the optimization model. PV systems include inverters that provide both active and reactive power capability for voltage support.


\textbf{Grid Connection} provides the interface to the main utility grid, enabling unidirectional active power flow but bidirectional reactive power flow, allowing the VPP to import active power when local generation is insufficient to meet demand while providing reactive power support for voltage regulation. The grid operates under time-varying electricity prices that incentivize optimal scheduling of DERs to minimize operational costs during normal grid-connected operation. During extreme weather events, the grid connection becomes unavailable, forcing the VPP into islanded operation where it must rely entirely on local resources. The system employs a binary weather state model where conditions are classified as either normal (grid available) or extreme weather (grid unavailable).




\textbf{The Control System} implements the MPC framework shown in Figure \ref{fig:vpp-arch}, which operates before, during, and after extreme weather events, seamlessly adapting its coordination strategy as conditions evolve. The system continuously processes time-varying demand and supply forecasts, electricity pricing data, weather state information, and real-time measurements from DERs to generate optimal control instructions for customer service prioritization, energy storage system dispatch, PV power distribution, and grid import scheduling. 

\textbf{The Error Simulation Model} incorporates realistic forecasting uncertainties into the system evaluation, simulating both systematic bias and horizon-dependent uncertainty for load demand and PV generation forecasts. The model enables assessment of the MPC framework's robustness under real-world conditions where perfect forecasts are unavailable, as detailed in Section \ref{sec:forecast_error}.

\subsection{Operational Phases}
\label{sec:timeline}
The MPC controller automatically adapts resource allocation based on current conditions and forecasts, operating across distinct temporal phases.

\textbf{Normal Grid-Connected Operation} focuses on economic optimization while maintaining power quality by ensuring bus voltages remain within operational limits. The controller leverages time-varying electricity prices to minimize operational costs through optimal scheduling of DERs. Battery systems perform energy arbitrage, storing energy during low-price periods and discharging during high-price periods.


\textbf{Pre-Event Preparation} begins when extreme weather first appears within the MPC prediction horizon (the controller's look-ahead window over which it optimizes at each time step), providing advance warning proportional to the horizon length. For example, with a 9-hour prediction horizon, extreme weather starting at hour 15 would trigger preparation activities beginning at hour 6. This relationship creates a fundamental trade-off: longer horizons provide more preparation time but suffer from reduced forecast accuracy, while shorter horizons offer more reliable predictions but less time to prepare. During this phase, battery charging strategies shift from peak shaving to ensuring adequate energy storage for islanded operation. The controller must balance preparation costs (charging batteries at potentially high prices) against the risk of insufficient energy during the outage, with decisions continuously updated as forecasts evolve and the event approaches.



\textbf{Islanded Operation} occurs during extreme weather events when the main grid becomes unavailable, forcing the VPP into self-sufficient operation. The system must rely entirely on local generation (PV systems) and stored energy (batteries) to maintain power delivery. Load shedding decisions prioritize customers based on their criticality levels (critical $>$ high $>$ low priority).


\textbf{Recovery Phase} follows extreme weather events as the main grid becomes available again. The VPP gradually transitions back to normal operation while continuing to serve previously disconnected loads. The control system maintains continuous operation throughout this transition, ensuring smooth restoration of services and optimal utilization of both grid power and local resources during the recovery period.


The multi-objective, multi-phase nature creates a complex optimization problem that requires the robust, adaptive coordination provided by the MPC framework detailed in the following sections.

\section{Mathematical Formulation}
\label{sec:math}
In this section, we present the complete mathematical formulation of the VPP coordination problem as a Mixed-Integer Quadratically Constrained Programming (MIQCP) optimization. The formulation includes operational and technical constraints (Section \ref{sec:constraints}), an enhanced delta optimization approach for power flow approximation (Section \ref{sec:delta}), and a multi-objective function (Section \ref{sec:objective}). Table \ref{tab:opt_variables} presents all optimization variables, while Table \ref{tab:parameters} provides system parameters, sets, and other notation used throughout the mathematical model.

\begin{table}[t]
\centering
\caption{Optimization Variables}
\label{tab:opt_variables}
\begin{tabular}{p{2cm}p{12cm}}
\toprule
\textbf{Variable} & \textbf{Description} \\
\midrule
$x^{t,c}_{\text{served}}$ & Binary variable indicating whether customer $c$ is served at time $t$ \\
$P^{t,p}_{\text{pv,self}}$ & Active power from PV system $p$ self-consumed by its owner at time $t$ \\
$P^{t,p}_{\text{pv,exp}}$ & Exported active power from PV system $p$ at time $t$ \\
$P^{t,p}_{\text{pv,curt}}$ & Curtailed active power from PV system $p$ at time $t$ \\
$Q^{t,p}_{\text{pv}}$ & Reactive power from PV system $p$ at time $t$ \\
$P^{t,b}_{\text{ch}}$ & Charging power of battery $b$ at time $t$ \\
$P^{t,b}_{\text{dch}}$ & Discharging power of battery $b$ at time $t$ \\
$\delta^{t,b}$ & Binary variable: 1 if battery $b$ is charging at time $t$, otherwise 0 \\
$P^{t,b}_{\text{batt}}$ & Net power of battery $b$ at time $t$  \\
$Q^{t,b}_{\text{batt}}$ & Reactive power of battery $b$ at time $t$ \\
$\text{SoC}^{t,b}$ & State-of-charge of battery $b$ at time $t$ \\
$P^{t}_{\text{grid}}$ & Active power imported from main grid at time $t$ \\
$Q^{t}_{\text{grid}}$ & Reactive power from main grid at time $t$ \\
$V^{t,n}$ & Voltage magnitude at bus $n$ at time $t$ \\
$P^{t,ij}_{\text{flow}}$ & Active power flow from bus $i$ to bus $j$ at time $t$ \\
$Q^{t,ij}_{\text{flow}}$ & Reactive power flow from bus $i$ to bus $j$ at time $t$  \\
$\Delta V^{t,n}$ & Change in squared voltage at bus $n$ between time $t-\Delta t$ and $t$ \\
$\Delta P^{t,ij}_{\text{flow}}$ & Change in active power flow from bus $i$ to bus $j$ between time $t-\Delta t$ and $t$  \\
$\Delta Q^{t,ij}_{\text{flow}}$ & Change in reactive power flow from bus $i$ to bus $j$ between time $t-\Delta t$ and $t$  \\
\bottomrule
\end{tabular}
\end{table}

\begin{table}[t]
\centering
\caption{System Parameters and Sets}
\label{tab:parameters}
\begin{tabular}{p{4.5cm}p{12cm}}
\toprule
\textbf{Symbol} & \textbf{Description} \\
\midrule
$\mathcal{C}$ & Set of all customers \\
$\mathcal{C}_n$ & Set of customers at bus $n$ \\
$\mathcal{P}$ & Set of all PV systems \\
$\mathcal{B}$ & Set of all batteries \\
$\mathcal{B}_n$ & Set of all batteries at bus $n$ \\
$\mathcal{L}$ & Set of all lines \\
$\Delta t$ & Time step  \\
$w_c\in\{w_{\text{critical}},w_{\text{high}},w_{\text{low}}\}$ & Priority weight of customer $c$ \\
$P^{t,c}_{\text{load}}$ & Forecasted load of customer $c$ at time $t$ \\
$P^{t,c}_{\text{load,actual}}$ & Actual load of customer $c$ at time $t$ \\
$P^{t,p}_{\text{pv}}$ & Forecasted active PV output for system $p$ at time $t$ \\
$P^{t,p}_{\text{pv,actual}}$ & Actual active PV output for system $p$ at time $t$ \\
$E^{\text{cap}}_{b}$ & Energy capacity of battery $b$ \\
$P^{\text{max}}_{b}$ & Maximum power rating of battery $b$  \\
$S^{\text{inv}}_{b}$ & Inverter capacity of battery $b$  \\
$\eta^{\text{ch}}_{t,b}$ & Charging efficiency at time $t$ of battery $b$ \\
$\eta^{\text{dch}}_{t,b}$ & Discharging efficiency at time $t$ of battery $b$ \\
$\text{SoC}^{\text{min}}_{b}$ & Minimum state-of-charge of battery $b$ \\
$\text{SoC}^{\text{max}}_{b}$ & Maximum state-of-charge of battery $b$ \\
$P^{\text{max}}_{p}$ & Maximum power rating of PV system $p$  \\
$S^{\text{inv}}_{p}$ & Inverter capacity of PV system $p$  \\
$r_{ij}, x_{ij}$ & Resistance and reactance of line $(i,j)$ \\
$V_{\text{min}}, V_{\text{max}}$ & Voltage magnitude limits \\
$\text{pf}_c$ & Power factor of customer $c$ \\
$\rho_t$ & Grid import price at time $t$ \\
$H$ & MPC prediction horizon \\
$V^{t,n}_{\text{ac}}$ & Voltage magnitude at bus $n$ at time $t$ from AC power flow analysis \\
$P^{t,ij}_{\text{ac}}$ & Active power flow from bus $i$ to bus $j$ at time $t$ from AC power flow analysis \\
$Q^{t,ij}_{\text{ac}}$ & Reactive power flow from bus $i$ to bus $j$ at time $t$ from AC power flow analysis \\
$\lambda_{\text{grid}}$ & Grid import cost weight coefficient \\
$\lambda_{\text{curt}}$ & PV curtailment penalty weight coefficient \\
$\lambda_{l}$ & Line losses penalty weight coefficient \\
$\lambda_{q}$ & Device reactive power penalty weight coefficient \\
\bottomrule
\end{tabular}
\end{table}

\subsection{Constraints}
\label{sec:constraints}
The optimization problem is subject to various operational and technical constraints that ensure feasible and safe system operation.

\paragraph{Battery Storage Constraints}
Battery operation is governed by mutually exclusive charging and discharging modes, enforced through binary variables. When $\delta^{t,b} = 1$, the battery charges with $P^{t,b}_{\text{ch}} \leq P^{\text{max}}_{b}$ and $P^{t,b}_{\text{dch}} = 0$. Conversely, when $\delta^{t,b} = 0$, the battery discharges with $P^{t,b}_{\text{dch}} \leq P^{\text{max}}_{b}$ and $P^{t,b}_{\text{ch}} = 0$. The net battery power is:
\begin{equation}
P^{t,b}_{\text{batt}} = P^{t,b}_{\text{ch}} - P^{t,b}_{\text{dch}}~.
\end{equation}

The battery state-of-charge evolves according to:
\begin{equation}
\text{SoC}^{t+\Delta t,b} = \text{SoC}^{t,b} + \left( \eta^{\text{ch}}_{t,b} \cdot P^{t,b}_{\text{ch}} - \frac{P^{t,b}_{\text{dch}}}{\eta^{\text{dch}}_{t,b}} \right) \frac{\Delta t}{E^{\text{cap}}_{b}}~.
\end{equation}

State-of-charge limits are enforced as $\text{SoC}^{\text{min}}_{b} \leq \text{SoC}^{t,b} \leq \text{SoC}^{\text{max}}_{b}$, and the inverter capacity constraint is
\begin{equation}
\label{eq:batt_inverter}
(P^{t,b}_{\text{batt}})^2 + (Q^{t,b}_{\text{batt}})^2 \leq (S^{\text{inv}}_{b})^2~.
\end{equation}

To prevent inefficient circular charging between batteries, the total battery charging power cannot exceed available generation:
\begin{equation}
\sum_{b \in \mathcal{B}} P^{t,b}_{\text{ch}} \leq P^{t}_{\text{grid}} + \sum_{p \in \mathcal{P}} P^{t,p}_{\text{pv,exp}}~.
\end{equation}

\paragraph{PV System Constraints}
The PV power balance ensures that total generation equals the sum of self-consumption, export, and curtailment:
\begin{equation}
P^{t,p}_{\text{pv,self}} + P^{t,p}_{\text{pv,exp}} + P^{t,p}_{\text{pv,curt}} = P^{t,p}_{\text{pv}}~.
\end{equation}

Self-consumed PV power is limited by the minimum of customer load and available PV generation, activated only when the customer is served\footnote{Since $P^{t,c}_{\text{load}}$ and $P^{t,p}_{\text{pv}}$ are exogenous forecasts, the right-hand side of Eq. \eqref{eq:min_pv} is $\text{constant}\times\text{binary}$ and adds no nonlinearity.}:
\begin{equation}
\label{eq:min_pv}
P^{t,p}_{\text{pv,self}} = \min(P^{t,c}_{\text{load}}, P^{t,p}_{\text{pv}}) \cdot x^{t,c}_{\text{served}}~.
\end{equation}

The PV inverter capacity constraint ensures the apparent power remains within equipment limits:
\begin{equation}
\label{eq:pv_inverter}
(P^{t,p}_{\text{pv,self}} + P^{t,p}_{\text{pv,exp}})^2 + (Q^{t,p}_{\text{pv}})^2 \leq (S^{\text{inv}}_{p})^2~.
\end{equation}

\paragraph{Power Balance Constraints}
Power balance constraints ensure energy conservation at each bus by requiring that the total power injected equals the total power withdrawn. 
\paragraph{Active Power Balance} The active power balance at bus $n$ is:
\begin{equation}
P^{t,n}_{\text{inj}} + \sum_{j:(j,n) \in \mathcal{L}} P^{t,jn}_{\text{flow}} = P^{t,n}_{\text{load}} + \sum_{j:(n,j) \in \mathcal{L}} P^{t,nj}_{\text{flow}}~.
\end{equation}

The net active power injection $P^{t,n}_{\text{inj}}$ includes contributions from PV exports and battery discharge:
\begin{equation}
P^{t,n}_{\text{inj}} = \sum_{c \in \mathcal{C}_n} P^{t,p_c}_{\text{pv,exp}} - \sum_{b \in \mathcal{B}_n} P^{t,b}_{\text{batt}}~.
\end{equation}

The total active load $P^{t,n}_{\text{load}}$ accounts for served customers and local PV self-consumption:
\begin{equation}
P^{t,n}_{\text{load}} = \sum_{c \in \mathcal{C}_n}\left(P^{t,c}_{\text{load}} \cdot x^{t,c}_{\text{served}} - P^{t,p_c}_{\text{pv,self}}\right)~,
\end{equation}
where $p_c$ denotes the PV system of customer $c$, and $P^{t,p_c}_{\text{pv,self}} = 0$ for customers without PV systems.

\paragraph{Reactive Power Balance} The reactive power balance follows the same structure:
\begin{equation}
Q^{t,n}_{\text{inj}} + \sum_{j:(j,n) \in \mathcal{L}} Q^{t,jn}_{\text{flow}} = Q^{t,n}_{\text{load}} + \sum_{j:(n,j) \in \mathcal{L}} Q^{t,nj}_{\text{flow}}~.
\end{equation}
The reactive power components are:
\begin{align}
Q^{t,n}_{\text{inj}} &= \sum_{c \in \mathcal{C}_n} Q^{t,p_c}_{\text{pv}} + \sum_{b \in \mathcal{B}_n} Q^{t,b}_{\text{batt}} \\
Q^{t,n}_{\text{load}} &= \sum_{c \in \mathcal{C}_n} Q^{t,c}_{\text{load}}~,
\end{align}
where the reactive load for each customer is determined by their power factor:
\begin{equation}
Q^{t,c}_{\text{load}} = P^{t,c}_{\text{load}} \cdot \tan(\arccos(\text{pf}_c)) \cdot x^{t,c}_{\text{served}}~.
\end{equation}

\paragraph{Grid Connection Constraints}
During extreme weather events, the main grid becomes unavailable:
\begin{equation}
P^{t}_{\text{grid}} = 0 \quad \text{and} \quad Q^{t}_{\text{grid}} = 0 \quad \text{if extreme weather at time } t.
\end{equation}

\paragraph{Voltage Constraints}
The voltage magnitude at each bus must remain within operational limits to ensure power quality:
$V_{\text{min}} \leq V^{t,n}\leq V_{\text{max}}$.

\subsection{Power flow approximation}
\label{sec:delta}
For radial distribution feeders, the DistFlow equation describes the fundamental voltage-power relationship across each branch \cite{BaranWu89}:
\begin{equation}
\label{eq:full-distflow}
 (V^{t,j}) ^2
= (V^{t,i}) ^2
 - 2\left(r_{ij} P^{t,ij}_{\text{flow}} + x_{ij} Q^{t,ij}_{\text{flow}}\right)
 + (r_{ij}^{2}+x_{ij}^{2})I_{t,ij}^{2}~,
\end{equation}
where $I_{t,ij}$ is the magnitude of the branch current.
To avoid complex nonlinearities that complicate the optimization, the standard LinDistFlow approximation neglects the rightmost term in Eq. \eqref{eq:full-distflow}, which represents losses due to the impedance magnitude \cite{9682997}. However, this loss term can become significant under heavy loading conditions or when operating far from nominal voltage levels, leading to substantial approximation errors that accumulate over the optimization horizon.

To address this limitation, we propose a delta optimization approach specifically tailored to the iterative nature of MPC. The key insight is that while absolute power flows and voltages may deviate significantly from nominal conditions, the changes between consecutive time steps are typically small, making linearization around the current operating point more accurate than linearization around a fixed nominal point.

Consider Eq. \eqref{eq:full-distflow} for the next time step $t+\Delta t$:
\begin{equation}
\label{eq:full-distflow1}
 (V^{t+\Delta t,j}) ^2
= (V^{t+\Delta t,i}) ^2
 - 2\left(r_{ij} P^{t+\Delta t,ij}_{\text{flow}} + x_{ij} Q^{t+\Delta t,ij}_{\text{flow}}\right)
 + (r_{ij}^{2}+x_{ij}^{2})I_{t+\Delta t,ij}^{2}~.
\end{equation}
We define the delta variables as:
\begin{align}
\Delta V^{t,n} &= (V^{t+\Delta t,n})^2-(V^{t,n})^2    \\
\Delta P^{t,ij}_{\text{flow}} &=    P^{t+\Delta t,ij}_{\text{flow}} -P^{t,ij}_{\text{flow}} \\
\Delta Q^{t,ij}_{\text{flow}} &=  Q^{t+\Delta t,ij}_{\text{flow}} -Q^{t,ij}_{\text{flow}}~.
\end{align}
Subtracting Eq. \eqref{eq:full-distflow} from Eq. \eqref{eq:full-distflow1} yields:
\begin{equation}
\label{eq:dist_flow_diff}
 \Delta V^{t,j}
= \Delta V^{t,i}
 - 2\left(r_{ij} \cdot \Delta P^{t,ij}_{\text{flow}} + x_{ij} \cdot \Delta Q^{t,ij}_{\text{flow}}\right)
 + (r_{ij}^{2}+x_{ij}^{2})\left(I_{t+\Delta t,ij}^{2}-I_{t,ij}^{2}\right)~.
\end{equation}

We approximate Eq. \eqref{eq:dist_flow_diff} by neglecting the rightmost term $(r_{ij}^{2}+x_{ij}^{2})\left(I_{t+\Delta t,ij}^{2}-I_{t,ij}^{2}\right)$. This approximation is justified because the term represents the difference between squared currents across consecutive time steps, which is typically much smaller in magnitude than the individual currents themselves. In addition, this difference approaches zero as the time step becomes smaller. Consequently, neglecting this differential loss term introduces significantly less error than neglecting the absolute loss term in the standard LinDistFlow approximation, particularly when the system operates under stressed conditions or far from nominal operating points.

At each MPC iteration, the voltage and power flow variables are initialized using the AC power flow results from the previous iteration:
\begin{align}
    V^{t,n} = V^{t-\Delta t,n}_{\text{ac}} \\ 
    P^{t,ij}_{\text{flow}} = P^{t-\Delta t,ij}_{\text{ac}} \\
    Q^{t,ij}_{\text{flow}} = Q^{t-\Delta t,ij}_{\text{ac}}~,
\end{align}
ensuring that each optimization begins from the actual system operating point (see Section \ref{sec:mpc}).
For the initial MPC iteration when no previous AC power flow results are available, all bus voltages are initialized to the slack bus voltage ($1.0$\,p.u.) and all power flows to zero.

\subsection{Objective Function for Optimization}
\label{sec:objective}
We employ a multi-objective function that balances three primary goals: maximizing system resilience through prioritized customer service, minimizing operational costs, and maximizing renewable energy utilization. In addition, we introduce two penalty terms to discourage unnecessary circulation of reactive power across the network.

\paragraph{Customer Service Maximization} The primary objective prioritizes power delivery based on customer criticality:
\begin{equation}
f_1(t) = \sum_{c \in \mathcal{C}} w_c \cdot x^{t,c}_{\text{served}}~,
\end{equation}
where $w_c$ represents the importance weight of customer $c$, and $x^{t,c}_{\text{served}}$ is a binary variable indicating service status. This formulation ensures critical loads (e.g., hospitals, emergency services) receive preferred treatment during resource-constrained conditions (extreme weather).

\paragraph{Grid Import Cost Minimization} The economic objective reduces operational expenses by minimizing electricity purchases from the main grid:
\begin{equation}
f_2(t) = \rho_t \cdot P^{t}_{\text{grid}} \cdot \Delta t~,
\end{equation}
where $\rho_t$ is the time-varying electricity price and $P^{t}_{\text{grid}}$ is the active power imported from the grid. 
This term incentivizes utilizing local generation and discharging batteries during high-price periods while charging batteries during low-price periods.

\paragraph{Renewable Energy Utilization} The sustainability objective maximizes clean energy usage by minimizing PV curtailment:
\begin{equation}
f_3(t) = \sum_{p \in \mathcal{P}} P^{t,p}_{\text{pv,curt}}~.
\end{equation}
This term encourages maximum utilization of available renewable generation, reducing the need for grid imports and battery discharge. This term is particularly important in the MPC framework, as it encourages storing renewable energy in batteries beyond what appears optimal within the limited prediction horizon, ensuring energy availability for future time steps that lie outside the current optimization window.

\paragraph{Reactive Power Management} 
Because the optimization problem does not inherently penalize reactive-power circulation, it can admit families of equivalent optimal solutions. The optimizer may therefore route reactive power between neighboring buses or co-located devices in ways that are cost-neutral yet physically wasteful.
To break this degeneracy and favor physically plausible operating points, we add two small quadratic regularizers (on branch flows and on device reactive power).

We define:
\begin{equation}
f_4(t) = \sum_{(i,j) \in \mathcal{L}} r_{ij} \left[ \left(P^{t,ij}_{\text{flow}}\right)^2 + \left(Q^{t,ij}_{\text{flow}}\right)^2 \right]~,
\end{equation}
which penalizes squared branch flows as a convex proxy for ohmic losses and discourages circulating flows.

For device-level regularization we use
\begin{equation}
f_5(t) = \sum_{p \in \mathcal{P}} \left(\frac{Q^{t,p}_{\text{pv}}}{S^{\text{inv}}_{p}}\right)^2 + \sum_{b \in \mathcal{B}} \left(\frac{Q^{t,b}_{\text{batt}}}{S^{\text{inv}}_{b}}\right)^2~,
\end{equation}
which penalizes unnecessary reactive injection/absorption at devices. By normalizing by inverter capacity, the penalty guides the optimizer to utilize larger inverters for reactive power needs, since they operate at lower utilization fractions and experience less heating and losses, reducing thermal stress on smaller inverters.

\paragraph{Complete Objective Function} The complete objective function, which we seek to maximize, combines all components:
\begin{equation}
F = \sum_{t=t_0}^{t_0+H} \Bigg[ f_1(t) - \lambda_{\text{grid}} f_2(t) - \lambda_{\text{curt}} f_3(t) - \lambda_l f_4(t) - \lambda_q f_5(t) \Bigg]~,
\label{objective_function}
\end{equation}
where $t_0$ represents the current time and $H$ is the prediction horizon in time units (e.g., hours). The summation advances through discrete time steps of duration $\Delta t$, covering the interval from $t_0$ to $t_0+H$.

The weighting coefficients ($\lambda_{\text{grid}}$, $\lambda_{\text{curt}}$, $\lambda_l$, $\lambda_q$) serve dual purposes: they perform necessary unit conversions to create a dimensionally consistent objective function, and they encode operational priorities. For instance, $\lambda_{\text{grid}}$ converts the cost term to the same units as the dimensionless customer service term, while simultaneously reflecting the relative importance of cost minimization versus service maximization.
The reactive power penalty weights $\lambda_l$ and $\lambda_q$ are set to very small values and act as tie-breakers rather than primary drivers, ensuring they promote efficient power quality management without dominating the main objectives.

The resulting MIQCP optimization problem defined by the objective function in Eq. \eqref{objective_function} subject to the constraints described in Sections \ref{sec:constraints} and \ref{sec:delta} is solved using the Gurobi optimizer.

\section{MPC vs Full-Horizon optimization}
\label{sec:mpc_vs_full}
In this section, we compare two different approaches to VPP coordination: the traditional full-horizon optimization and Model Predictive Control with receding horizon optimization. Both methods solve the same underlying optimization problem defined in Eq. \eqref{objective_function}, but differ significantly in their implementation strategy and robustness to real-world uncertainties.

\subsection{Full-Horizon Optimization Approach}
The full-horizon optimization approach, illustrated in Figure \ref{fig:flowchart_full}, represents the traditional method for VPP coordination where the entire control problem is solved in a single comprehensive optimization. This method begins by obtaining forecasts for customer loads, PV generation output, energy import costs, and weather conditions over the complete planning period (e.g. 24-48 hours). The optimization problem defined in Eq. \eqref{objective_function} is then solved once with $t_0=0$ and $H$ covering the entire planning period, producing a comprehensive resource allocation plan that specifies control strategies for all time steps simultaneously.

While this approach appears optimal from a mathematical standpoint, its practical implementation reveals significant challenges when forecast errors and modeling uncertainties are present.
As the pre-computed plan is executed sequentially, at each time step the system obtains actual measurements, including real customer loads, actual PV generation outputs, and updated battery parameters based on measured temperatures. 
To address feasibility violations that arise from discrepancies between these actual values and the original forecasts, the full-horizon approach incorporates verification and corrective mechanisms. Before implementing any control action, the system performs three feasibility checks:
\begin{itemize}
    \item \textbf{Battery Dispatch Feasibility:} Violations occur when forecasting errors in load or PV output affect planned charging/discharging schedules, or when temperature variations alter battery efficiency parameters that were not modeled in the original optimization. The system verifies that planned active and reactive power dispatch satisfies inverter capacity constraints (Eq. \eqref{eq:batt_inverter}) and can be delivered given current SoC, efficiency parameters, and charging/discharging limits.
    \item \textbf{PV Power Distribution Feasibility:} Violations arise when actual generation or customer demand differs from forecasted values, making planned self-consumption levels impossible to achieve. The system validates that power allocation among self-consumption, export, and curtailment equals actual PV output, respects customer load limits, and satisfies inverter capacity constraints (Eq. \eqref{eq:pv_inverter}).
    \item \textbf{Grid Import Feasibility:} Violations occur during extreme weather events when load forecasting errors result in insufficient battery pre-charging and the system subsequently requires grid power that is unavailable during outages. Using full AC power flow analysis, the system determines the required grid import and flags infeasibility when positive import is needed during grid outages.
\end{itemize}

When violations are detected, corrective re-optimization is performed for the current time step only, optimizing Eq. \eqref{objective_function} with $t_0$ set to the current time and $H = 0$, using actual rather than forecasted values. While this corrective mechanism ensures feasibility, it fundamentally abandons the globally optimized strategy in favor of locally feasible but potentially suboptimal decisions.

\subsection{Model Predictive Control Approach}
\label{sec:mpc}

\begin{figure}
    \centering
    \begin{subfigure}[b]{0.49\linewidth}
        \centering
        \includegraphics[width=\linewidth]{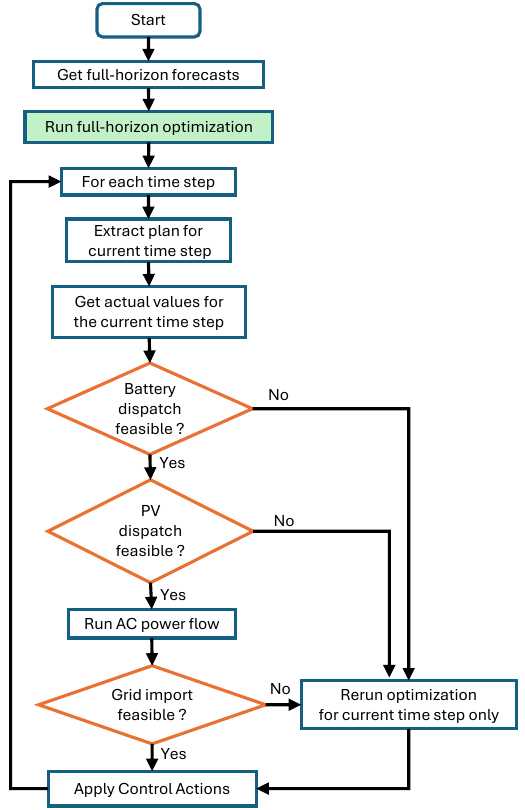}
        \caption{Full-Horizon Optimization}
        \label{fig:flowchart_full}
    \end{subfigure}
    \hfill
    \begin{subfigure}[b]{0.44\linewidth}
        \centering
        \includegraphics[width=0.8\linewidth]{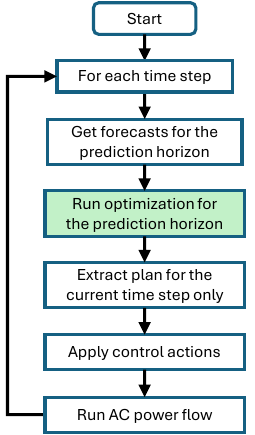}
        \caption{Model Predictive Control}
        \label{fig:flowchart_mpc}
    \end{subfigure}
    \caption{VPP control strategies. (a) Full-Horizon optimization: a single plan is computed for the entire simulation period, with feasibility checks and corrective re-optimization when violations occur. (b) Model predictive control (MPC): a rolling-horizon optimization is solved at each time step, and only the current control actions are implemented before forecasts and states are updated.}
    \label{fig:control_strategies}
\end{figure}

The MPC framework, shown in Figure \ref{fig:flowchart_mpc}, addresses the limitations of full-horizon optimization through a receding-horizon strategy that continuously adapts to evolving system conditions. Rather than committing to a fixed long-term plan, MPC operates through repeated optimization cycles, each covering a limited prediction horizon (e.g. 4-10 hours). At each time step, the controller obtains updated forecasts for customer loads, PV generation, energy costs, and weather conditions, incorporating the most recent available information and system measurements.
The optimization problem in Eq. \eqref{objective_function} is then solved with $t_0$ representing the current time step and $H$ denoting the MPC prediction horizon. This produces optimal resource allocation decisions over the prediction window. Crucially, only the control decisions for the immediate (current) time step are extracted and implemented, whereas all future decisions are ignored.
The controller then advances the time horizon by setting $t_0:=t_0+\Delta t$ and repeats the entire optimization process, creating the receding horizon behavior where the prediction window continuously slides forward in time.

The iterative nature of MPC benefits from the delta optimization approach described in Section \ref{sec:delta}. At each iteration, after applying the optimized control actions, a full AC power flow analysis provides accurate reference values $V^{t,n}_{\text{ac}}$,$P^{t,ij}_{\text{ac}}$, $Q^{t,ij}_{\text{ac}}$ that represent the true operating point. These values serve as the linearization center for the subsequent MPC iteration, preventing accumulation of linearization errors and naturally incorporating system feedback.
The comparative performance of MPC vs. full-horizon under various uncertainty scenarios is evaluated in Section \ref{sec:result_analysis}.

\section{Forecasting Uncertainties and Model Parameters Evolution}
\label{uncertain}
To evaluate our proposed MPC approach under realistic conditions, we implement explicit models for forecasting errors and time-varying system parameters. 
In this work, we assume that energy costs and weather classification (normal vs. extreme weather) are perfectly forecasted.
In Section \ref{sec:forecast_error} we present forecast error models for load demand and PV generation. Then, in Section \ref{sec:battery} we describe the time-varying battery efficiency model that captures thermal effects on charging and discharging performance.

\subsection{Forecast Error Models}
\label{sec:forecast_error}
We model forecasting errors to reflect two realistic characteristics: systematic bias and increasing uncertainty with prediction horizon. Systematic bias captures consistent forecasting tendencies, such as load forecasts that regularly underestimate peak demand or PV forecasts that overestimate generation during cloudy conditions. The uncertainty component increases with prediction horizon, reflecting the fundamental principle that short-term forecasts are inherently more accurate than long-term forecasts.

These forecast errors are applied during the MPC forecasting process: at each optimization cycle, forecasts for future time steps ($t>t_0$) incorporate both bias and uncertainty according to the models below, while the current time step forecasts ($t=t_0$) equal the actual values. For full-horizon optimization, forecast errors are applied to the initial forecasts for the entire planning period, and these forecast deviations persist throughout execution since the pre-computed plan is not updated with new information.

\paragraph{Load Forecast Errors}
Load forecasts are modeled as:
\begin{equation}
P^{t,c}_{\text{load}} = P^{t,c}_{\text{load,actual}} \cdot (1 + b_{\text{load}} + \epsilon^{t,c}_{\text{load}})~,
\end{equation}
where $b_{\text{load}}$ is a systematic bias term (positive for over-forecasting, negative for under-forecasting) and $\epsilon^{t,c}_{\text{load}} \sim \mathcal{N}(0, \sigma^2_{\text{load}}(t))$ represents random forecast error with standard deviation:
\begin{equation}
\sigma_{\text{load}}(t) = \alpha_{\text{load}} \cdot (t - t_0)~.
\end{equation}
The parameter $\alpha_{\text{load}}$ controls how rapidly forecast uncertainty degrades with prediction horizon. Forecasted loads are constrained to be non-negative: $P^{t,c}_{\text{load}} = \max(P^{t,c}_{\text{load}}, 0)$.

\paragraph{PV Generation Forecast Errors}
PV forecasts are modeled similarly:
\begin{equation}
P^{t,p}_{\text{pv}} = P^{t,p}_{\text{pv,actual}} \cdot (1 + b_{\text{pv}} + \epsilon^{t,p}_{\text{pv}})~,
\end{equation}
where $\epsilon^{t,p}_{\text{pv}} \sim \mathcal{N}(0, \sigma^2_{\text{pv}}(t))$ with $\sigma_{\text{pv}}(t) = \alpha_{\text{pv}} \cdot (t - t_0)$.
To ensure physical realizability, we clamp: 
$P^{t,p}_{\text{pv}} = \min(\max(P^{t,p}_{\text{pv}}, 0), P_p^{\max})$.

\subsection{Time-varying battery efficiency model}
\label{sec:battery}
To capture internal parameter evolution, we implement a temperature-dependent battery efficiency model that updates between MPC iterations based on thermal dynamics. We use fixed efficiency values within each optimization cycle for computational tractability, then simulate thermal behavior using the optimized power set-points to compute updated efficiencies for the subsequent cycle.

The electro-thermal model captures three heat generation mechanisms: charging conversion losses $P^{t}_{\text{ch}}(1-\eta^{\text{ch}}_{t})$, discharging conversion losses $P^{t}_{\text{dch}}(1/\eta^{\text{dch}}_{t} - 1)$, and Joule heating from temperature-dependent internal resistance \cite{FORGEZ20102961}:
\begin{equation}
R_{\text{int},T} = R_{\text{int},0}\cdot \left(1 + \beta(T_{\text{ref}} - T)\right)~,
\end{equation}
where $R_{\text{int},0}$ is the internal resistance at reference temperature $T_{\text{ref}}$, $\beta$ is the temperature coefficient, and $T$ represents cell temperature.

Total heat generation combines all loss mechanisms \cite{10403325}:
\begin{equation}
Q_t = P^{t}_{\text{ch}}\bigg(1 - \eta^{\text{ch}}_{t}\bigg) + P^{t}_{\text{dch}}\bigg(\frac{1}{\eta^{\text{dch}}_{t}} - 1\bigg) + \frac{(P^{t}_{\text{batt}})^2R_{\text{int},T_t}}{(U^{\text{batt}})^2}~,
\end{equation}
where $U^{\text{batt}}$ is the battery terminal voltage.

Cell temperature evolves according to first-order thermal dynamics \cite{FORGEZ20102961}:
\begin{equation}
T_{t+\Delta t} = T_t + \bigg(Q_t - \frac{T_t - T_t^{\text{amb}}}{R_{\theta}}\bigg)\frac{\Delta t}{C_{\theta}}~,
\end{equation}
where $R_{\theta}$ is thermal resistance, $C_{\theta}$ is thermal capacity, and $T_t^{\text{amb}}$ is ambient temperature.  


Efficiencies follow quadratic temperature relationships \cite{10403325, OLLAS2023108048}:
\begin{align}
\eta^{\text{ch}}_{t+\Delta t}&= \eta^{\text{ch}}_{0}\bigg[1 - \alpha_c(T_{t+\Delta t} - T_{\text{ref}})^2\bigg] \\
\eta^{\text{dch}}_{t+\Delta t} &= \eta^{\text{dch}}_{0}\bigg[1 - \alpha_d(T_{t+\Delta t} - T_{\text{ref}})^2\bigg]~,
\end{align}        
where $\eta^{\text{ch}}_{0}$ and $\eta^{\text{dch}}_{0}$ are nominal efficiencies at $T_{\text{ref}}$, and $\alpha_c$, $\alpha_d$ are temperature coefficients characterizing efficiency degradation rates. These updated efficiency values allow the MPC controller to adapt to thermal effects in the subsequent optimization cycle.

\section{Modeling approach}
\label{sec:simulations}

The proposed MPC framework is general and applicable to any extreme weather scenario (e.g., hurricanes, ice storms, floods, wildfires) causing grid outages and to any radial distribution topology. We validate the approach using a UK heatwave case study on the IEEE 33-bus radial distribution network, a widely adopted benchmark system for distribution network studies.

\subsection{Input Data}
\label{sec:sim-data}

A 24-hour horizon on 19~July~2022 (UK record heat day; 40.3\,$^{\circ}$C at Coningsby \cite{metoffice2022record}) is simulated with hourly import prices from Elexon BMRS Market Index Prices (MIP) \cite{bmrs_mip_page}. A temperature threshold of $27^{\circ}$C defines the heatwave. The extreme-weather window, during which grid import is unavailable, is fixed at 12{:}00–18{:}00~UTC. The scenario configuration and exogenous inputs are listed in Table~\ref{tab:scenario-inputs}. The network asset inventory and aggregate statistics used by the controller and optimization model are reported in Table~\ref{tab:assets-summary}.

\begin{table}[t]
\centering
\caption{Scenario and exogenous inputs (19~July~2022, UTC)}
\label{tab:scenario-inputs}
\begin{tabular}{l l}
\toprule
Simulation horizon              & 00{:}00--23{:}00 (1\,h step) \\
Weather day                     & UK heatwave; threshold $27^{\circ}$C \\
Extreme-weather window          & 12{:}00--18{:}00 \\
Import price signal             & Elexon BMRS MIP (hourly) \\
Avg / Min / Max price           & 0.36 / 0.20 / 0.54 GBP/kWh \\
Price min / max time            & 23{:}00 / 18{:}00 \\
\bottomrule
\end{tabular}
\end{table}

\begin{table}[t]
\centering
\caption{Network assets and aggregate statistics}
\label{tab:assets-summary}
\begin{tabular}{l r}
\toprule
Load buses (bus~1 slack)                           & 32 \\
Priority counts (Critical / High / Low)            & 12 / 10 / 10 \\
Power factor (min / mean / max)                    & 0.90 / 0.954 / 0.98 \\
Daily load energy                                   & 86.44\,MWh \\
Peak load (time)                                    & 4.068\,MW @ 17{:}00 \\
\midrule
Buses with PV                                       & 28 \\
Total PV capacity                                   & 1.857\,MW \\
Per-bus PV capacity (min / mean / max)             & 24.77 / 66.34 / 92.87\,kW \\
PV inverter ratings                          & 30 / 40 / 100\,kVA \\
Daily PV energy / share                             & 15.08\,MWh / 17.4\% \\
Peak PV (time)                                      & 1.857\,MW @ 11{:}00 \\
\midrule
Buses with ESS                                      & 32 \\
Total ESS energy / power                            & 2.542\,MWh / 1.272\,MW \\
Per-unit ESS energy (min / mean / max)             & 38.59 / 79.44 / 165.5\,kWh \\
Per-unit ESS power (min / mean / max)              & 19.3 / 39.74 / 82.8\,kW \\
ESS inverter ratings                         & 25 / 45 / 70 / 90\,kVA \\
Initial SoC                             & 0.15 \\
SoC bounds                             & [0.15, 0.85] \\
\midrule
Peak net demand (time)                              & 3.786\,MW @ 18{:}00 \\
\bottomrule
\end{tabular}
\end{table}


\subsection{Parameter Selection}
The simulation employs a temporal discretization of $\Delta t=1$\,hour.

\paragraph{Voltage Limits}
Following the GB Security and Quality of Supply Standard (SQSS), which specifies a minimum voltage of $0.94$\,p.u., we set $V_{\text{min}} = 0.947$\,p.u. in the optimization to provide headroom, since the actual voltage calculated by AC power flow is typically lower than the linearized approximation used in the optimization. Symmetrically, we set $V_{\text{max}} = 1.053$\,p.u.

\paragraph{Customer Priority Weights}
Priority weights are normalized with the critical weight as unity: $w_{\text{critical}} = 1, w_{\text{high}} = 0.1, w_{\text{low}} = 0.02$. This relative weighting ensures critical customers receive strong preference during load shedding.

\paragraph{Battery Parameters}
All batteries operate within SoC bounds  $\text{SoC}^{\text{min}}= 0.15$ and $\text{SoC}^{\text{max}} = 0.85$. The charging and discharging efficiencies are fixed at $\eta^{\text{ch}} = \eta^{\text{dch}}=0.95$ for all experiments except the time-varying battery efficiency case presented in Section \ref{sec:sim_eff}.

\paragraph{Battery Electro–Thermal Parameters}
All batteries share the same nominal electrical parameters at \(T_{\text{ref}}=25^{\circ}\mathrm{C}\).
Thermal parameters were calibrated zone-wise from step–power tests. Within each MPC cycle the efficiencies are fixed and updated between cycles by the thermal model in Section~\ref{sec:battery}.
Table~\ref{tab:batt_therm_params} reports the symbols, units, and values used in this study.

\begin{table}[t]
\centering
\caption{Battery electro–thermal efficiency parameters used in simulations (values at \(T_{\text{ref}}=25^{\circ}\mathrm{C}\)).}
\label{tab:batt_therm_params}
\begin{tabular}{l l l}
\toprule
\textbf{Symbol / Item} & \textbf{Meaning / Unit} & \textbf{Value}\\
\midrule
$\alpha_c$ & Quad. temp. coeff. (charge) [$^{\circ}\mathrm{C}^{-2}$] & $10^{-4}$  \\
$\alpha_d$ & Quad. temp. coeff. (discharge) [$^{\circ}\mathrm{C}^{-2}$] & $10^{-4}$  \\
$R_{\text{int},0}$ & Internal resistance at $T_{\text{ref}}$ [\(\Omega\)] & $0.0035$  \\
$\beta$ & Temp. coeff. of $R_{\text{int}}$ [$^{\circ}\mathrm{C}^{-1}$] & $0.003$  \\
$R_{\theta}$ & Thermal resistance [$^{\circ}\mathrm{C}/\mathrm{kW}$] & $[2.771,\;4.486]$  \\
$C_{\theta}$ & Thermal capacity [kWh/$^{\circ}\mathrm{C}$] & $[0.104,\;0.439]$  \\
$U^{\text{batt}}$ & Terminal voltage [V] & $900$  \\
$T_{\text{ref}}$ & Reference temperature [$^{\circ}\mathrm{C}$] & $25.0$ \\
$T^{\text{amb}}_0$ & Initial ambient temperature [$^{\circ}\mathrm{C}$] & $22.0$ \\
\bottomrule
\end{tabular}

\vspace{3pt}
\end{table}

\paragraph{Objective Function Weights}
The grid import cost weight $\lambda_{\text{grid}} = 10^{-6}$ is selected to ensure that even low priority customers are served during \textit{normal} weather conditions. Specifically, this value satisfies $\lambda_{\text{grid}} < w_{\text{low}}/(\max_t \rho_t  \cdot \max_{t,c} P^{t,c}_{\text{load}} \cdot \Delta t)$, preventing the cost term from dominating customer service objectives. The PV curtailment penalty $\lambda_{\text{curt}}=0.1$ was empirically determined to achieve negligible curtailment. The reactive power penalties $\lambda_l= 10^{-14}$ and $\lambda_q=10^{-15}$ are set to extremely small values, acting purely as tie-breakers to discourage unnecessary reactive power circulation without influencing the primary optimization objectives.

\subsection{Performance Metrics}
We focus our experimental analysis on two primary performance metrics:

\paragraph{Resilience metric}
We define a resilience metric that quantifies the ability of the VPP to maintain power delivery to priority customers during extreme weather events. The metric is calculated as the ratio of the actual weighted customer service achieved during the outage period to the maximum possible weighted service if all customers were served throughout the event. Specifically, for the extreme weather duration spanning time steps $T_{\text{extreme}}$, the resilience metric is:
\begin{equation}
R = \frac{\sum_{t \in T_{\text{extreme}}} \sum_{c \in C} w_c \cdot x^{t,c}_{\text{served}}}{\sum_{t \in T_{\text{extreme}}} \sum_{c \in C} w_c}
\end{equation}
where recall that $w_c$ denotes the priority weight of customer $c$ and $x^{t,c}_{\text{served}}$ is the binary service status. This metric ranges from 0 (complete failure) to 1 (full service), and inherently emphasizes the importance of maintaining service to critical customers. For instance, serving all critical customers but no low-priority customers would yield a higher resilience score than serving all low-priority customers but losing critical loads, reflecting the operational reality that not all customers have equal societal impact during emergencies. This metric enables quantitative comparison between different control strategies and provides a single interpretable measure of system resilience performance.

\paragraph{Total operational cost}
The total operational cost quantifies the economic efficiency of VPP coordination. It is calculated as the sum of electricity purchases from the grid across all time steps:
\begin{equation}
C_{\text{total}} = \sum_t \rho_t \cdot P_{\text{grid}}^t \cdot \Delta t
\end{equation}
where $\rho_t$ is the time-varying electricity price (GBP/kWh) and $P_{\text{grid}}^t$ is the active power imported from the grid at time $t$. 
Lower costs indicate efficient exploitation of electricity price variations and local resource utilization, though this must be balanced against resilience requirements. The key tension arises during the pre-event phase: achieving high resilience requires aggressive battery pre-charging, which may occur during high-price periods, thus increasing operational costs.

\section{Results and Analysis}
\label{sec:result_analysis}

\subsection{Baseline Case: Ideal Conditions}

\begin{figure}[t!]
    \centering
    \begin{subfigure}[b]{0.49\textwidth}
    \centering
    \includegraphics[width=\textwidth]{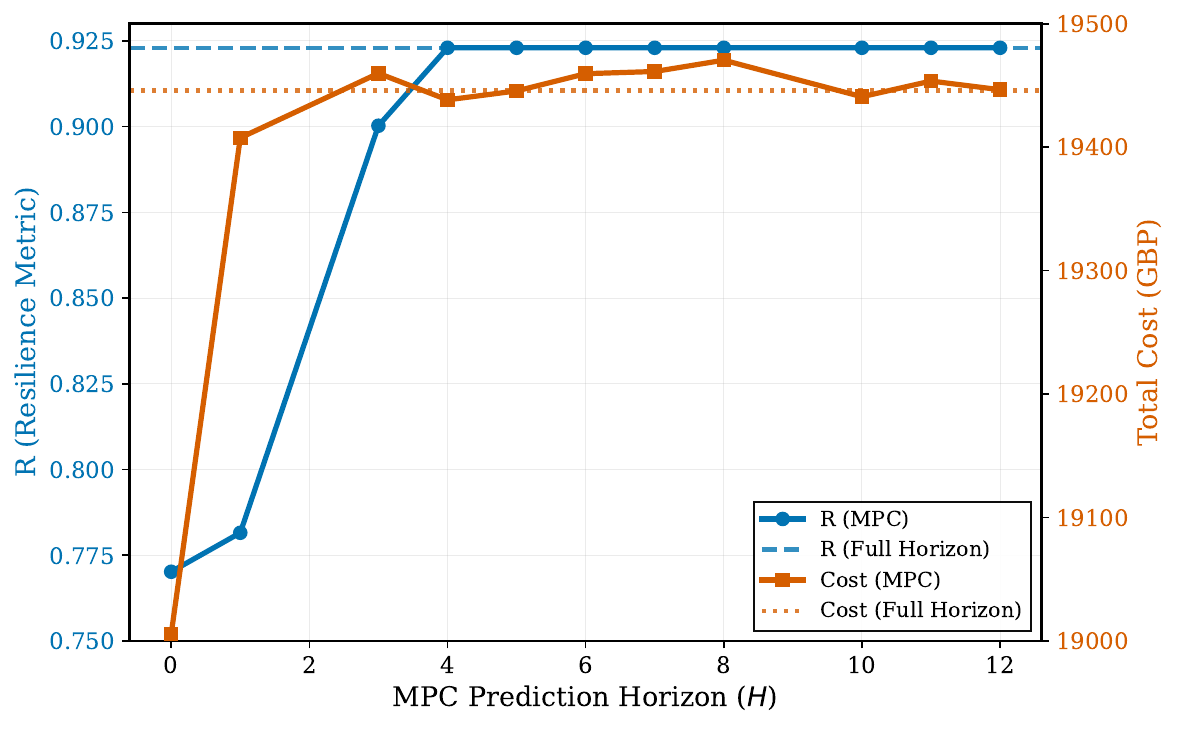}
    \caption{MPC Prediction Horizon effect}
    \label{fig:mpc_prediction_horizon}
    \end{subfigure}
    \hfill
    \begin{subfigure}[b]{0.49\textwidth}
    \centering 
    \includegraphics[width=\textwidth]{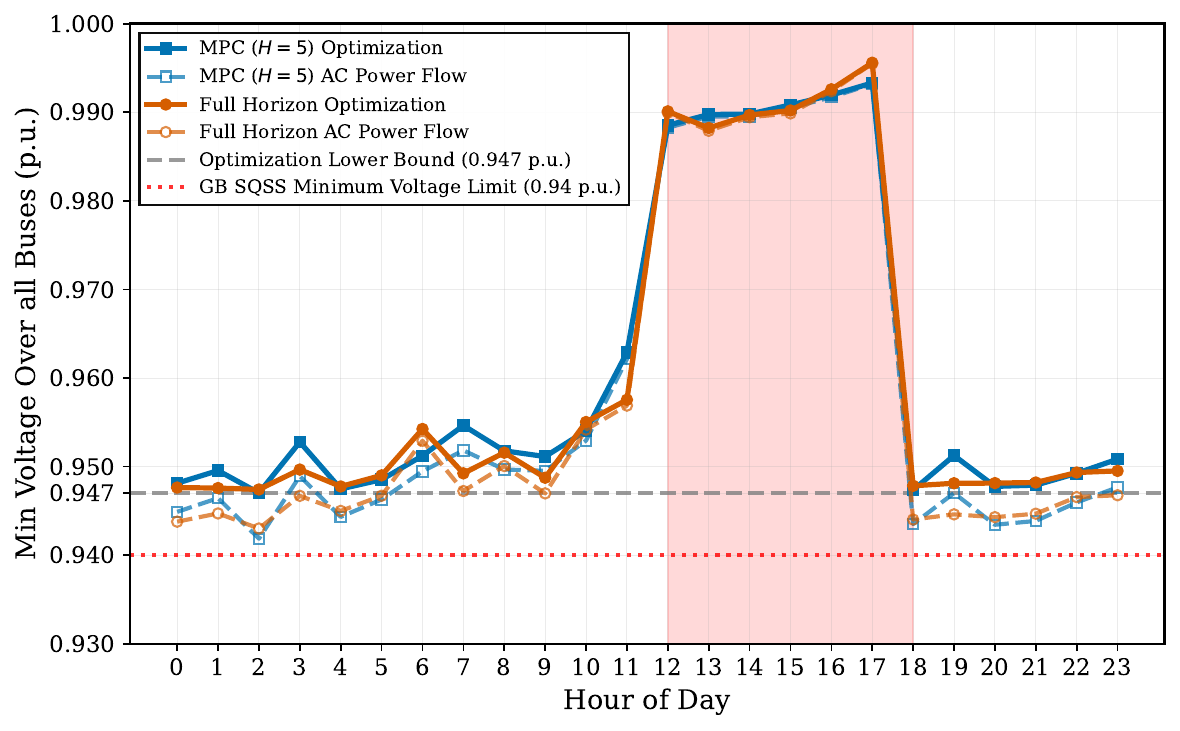}
    \caption{Voltage comparison}
    \label{fig:voltage}
    \end{subfigure}
    \hfill
    \begin{subfigure}[b]{0.49\textwidth}
    \centering 
    \includegraphics[width=\textwidth]{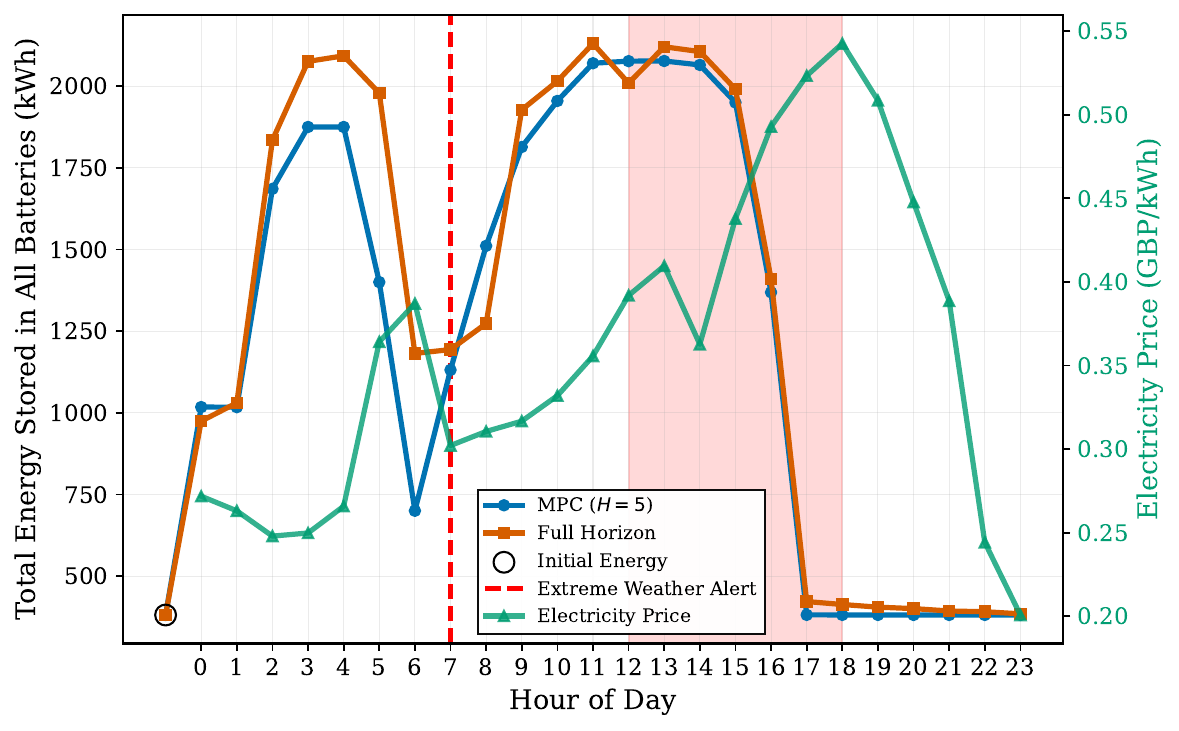}
    \caption{Total battery energy and electricity price}
    \label{fig:battery_energy}
    \end{subfigure}
    \hfill
    \begin{subfigure}[b]{0.49\textwidth}
    \centering 
    \includegraphics[width=\textwidth]{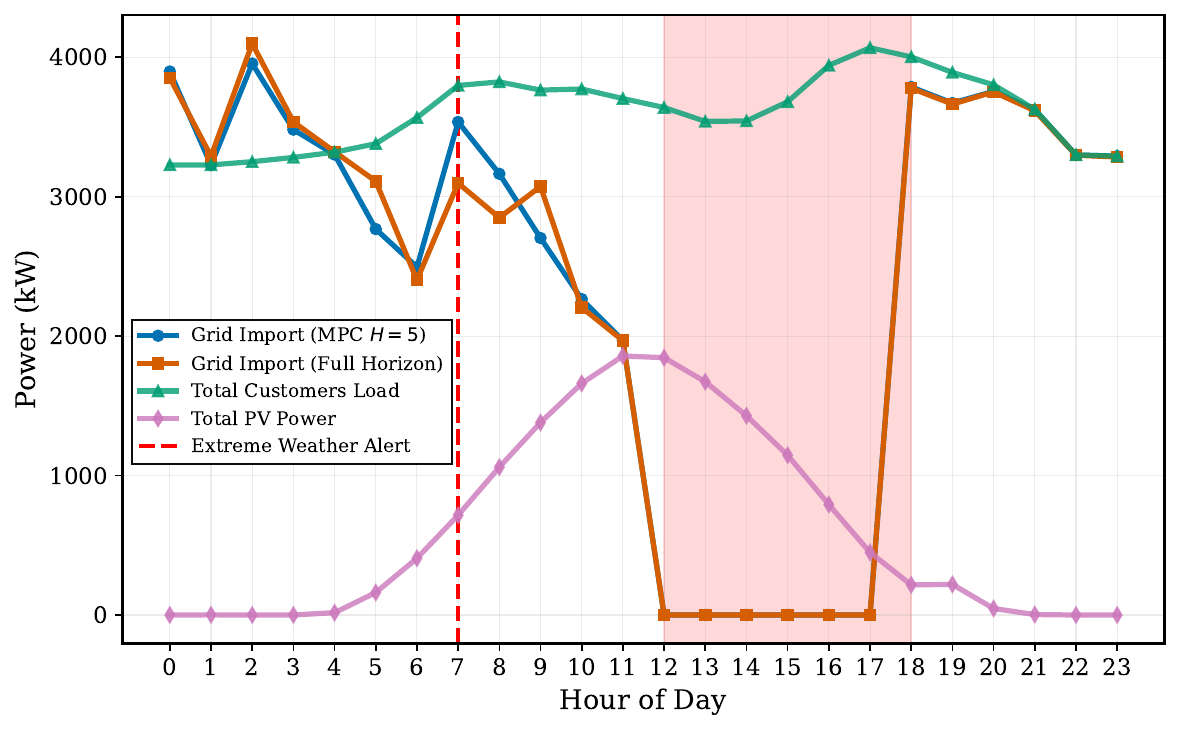}
    \caption{Grid import, Total load, Total PV}
    \label{fig:grid_import}
    \end{subfigure}
    \caption{MPC vs Full-Horizon under ideal conditions.}
    \label{fig:1}
\end{figure}

We first evaluate the system assuming perfect forecasts and no modeling uncertainties (i.e. fixed battery efficiencies). Under these ideal conditions, full-horizon optimization represents the theoretical global optimum, providing an upper bound on achievable performance.

\paragraph{MPC Prediction Horizon Selection}
Figure \ref{fig:mpc_prediction_horizon} illustrates the relationship between MPC prediction horizon ($H$) and system performance, measured through both resilience ($R$) and total cost. For short prediction horizons ($H < 4$\,hours), the MPC exhibits degraded resilience performance because the controller lacks sufficient look-ahead capability to anticipate and prepare for the extreme weather event. As the prediction horizon increases beyond 4\,hours, resilience performance plateaus and matches the full-horizon performance. Simultaneously, total cost increases as the MPC implements more aggressive battery pre-charging strategies. 

Based on this analysis, we select $H = 5$ hours as the baseline MPC prediction horizon for subsequent experiments. This choice provides the minimum look-ahead capability needed to achieve optimal resilience while avoiding the computational overhead of longer horizons that yield diminishing returns.

\paragraph{Voltage Quality}
Figure \ref{fig:voltage} demonstrates that both MPC and full-horizon optimization successfully maintain voltage quality throughout the simulation period. The minimum voltage across all buses from both optimization solutions remains consistently above the hard constraint of 0.947\,p.u. imposed in the optimization formulation. More importantly, the actual voltages calculated by AC power flow analysis remain above the GB SQSS minimum voltage limit of 0.94\,p.u. (shown as the lower red dashed line).

Beyond meeting minimum voltage requirements, MPC achieves slightly better overall voltage quality compared to full-horizon optimization. The average of the minimum voltage across buses over the simulation period is 0.9607\,p.u. for MPC versus 0.9601\,p.u. for full-horizon, indicating that MPC maintains voltages marginally closer to nominal values throughout the day.

For the MPC approach, the difference between the optimization solution and the AC power flow analysis is consistently small, remaining below 0.005\,p.u. throughout the simulation period. This close agreement validates the accuracy of the delta optimization approach introduced in Section \ref{sec:delta}.

\paragraph{Battery Pre-Charging and Energy Arbitrage}
Figure \ref{fig:battery_energy} shows the coordinated battery management strategies alongside the electricity price profile. The total energy stored across all batteries exhibits distinct operational phases driven by the dual objectives of resilience preparation and economic optimization.

During the early morning hours (hours 0-2), batteries charge to exploit low electricity prices, with maximum charging occurring at hour 2 when prices reach a local minimum. Following this initial charging event, batteries partially discharge during hours 3-6 to serve morning demand and reduce grid imports during periods of moderate pricing and limited PV generation. Notably, at hour 6 when prices reach a local maximum, batteries discharge to minimize expensive grid imports, demonstrating economic arbitrage behavior during normal operation.

At hour 7 ($H = 5$ hours before the extreme weather event begins) the MPC controller first detects the impending outage within its prediction horizon. This moment, marked as "Extreme Weather Alert" in Figure \ref{fig:battery_energy}, triggers a shift in MPC's control strategy from economic optimization to resilience preparation. Full-horizon optimization also begins charging at hour 7, as this timing represents the optimal balance between preparation costs and resilience requirements in its global optimization. From hour 7 onwards, both MPC and full-horizon execute rapid battery charging, achieving near-maximum storage capacity by hour 11. This aggressive pre-charging occurs despite elevated electricity prices during hours 7-11, explicitly prioritizing resilience over short-term cost minimization. The near-identical charging trajectories between MPC and full-horizon confirm that the selected prediction horizon provides sufficient look-ahead for optimal preparation.

During the extreme weather period (hours 12-18), the discharge pattern reflects the utilization of available resources. Initially, during hours 12-15 when substantial PV generation remains available (see Figure \ref{fig:grid_import}), battery discharge is relatively modest as solar energy serves priority loads. As PV output declines toward sunset (hours 16-18), battery discharge accelerates to compensate for the diminishing renewable generation. By the end of the outage, total stored energy has decreased to approximately the minimum SOC constraints across the distributed battery fleet.

\paragraph{Grid Import and Power Balance}
Figure \ref{fig:grid_import} illustrates the temporal dynamics of grid import, total customer load, and aggregate PV generation throughout the simulation period. The grid import profile directly reflects the battery charging strategies discussed above, with maximum grid import at hour 2 corresponding to economic arbitrage during the lowest price period, and minimum grid import at hour 6 during the local price peak when batteries discharge to reduce expensive imports.

During daytime hours (6-18), PV generation significantly reduces net grid import requirements. The controller leverages this renewable generation both to serve local loads through self-consumption and to charge batteries through PV export. The extreme weather period (hours 12-18, red shaded region) demonstrates complete grid disconnection with zero import, forcing the system to rely entirely on battery discharge and residual PV generation to maintain partial load service. The gap between total customer load and available resources necessitates selective load shedding based on customer priority classifications, resulting in the resilience levels of $R \approx 0.92$ observed in Figure \ref{fig:mpc_prediction_horizon}.

Comparison between MPC and full-horizon shows nearly overlapping grid import profiles throughout the simulation. This close agreement under ideal conditions validates that MPC with an appropriate prediction horizon successfully replicates the globally optimal strategy.

\subsection{Impact of Load and PV Forecasting Errors}

\begin{figure}[t!]
    \centering
    \begin{subfigure}[b]{0.49\textwidth}
    \centering
    \includegraphics[width=\textwidth]{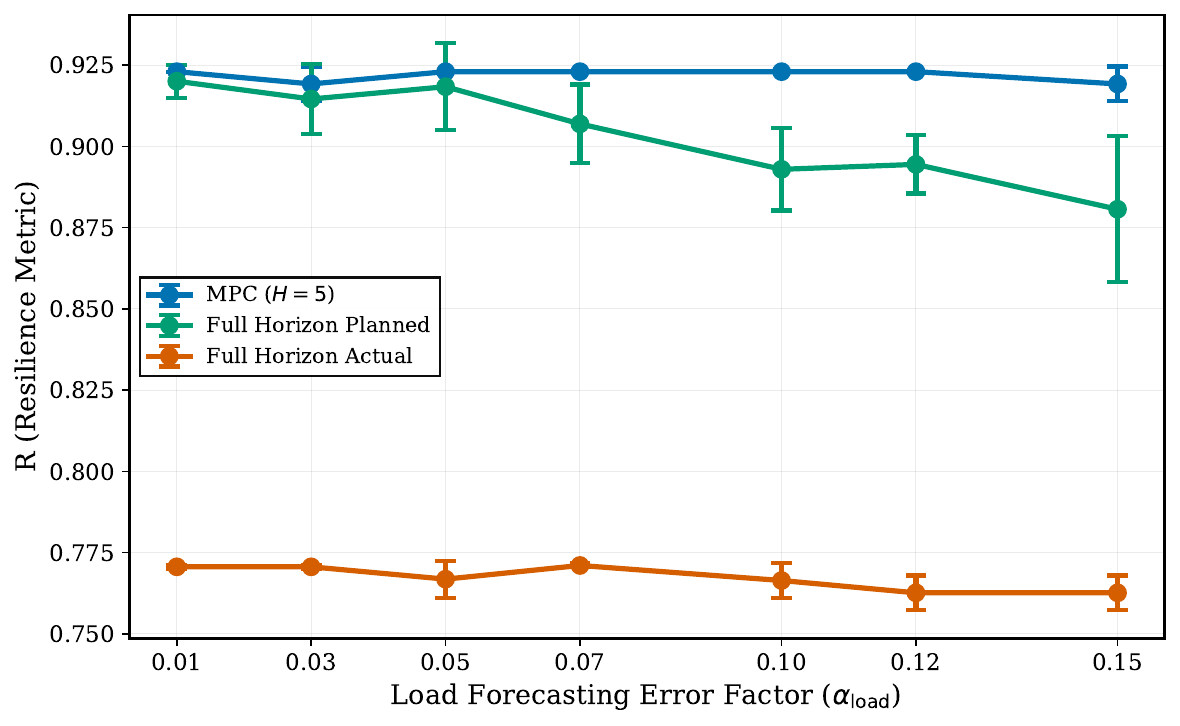}
    \caption{Effect of load forecasting error}
    \label{fig:load_error}
    \end{subfigure}
    \hfill
    \begin{subfigure}[b]{0.49\textwidth}
    \centering 
    \includegraphics[width=\textwidth]{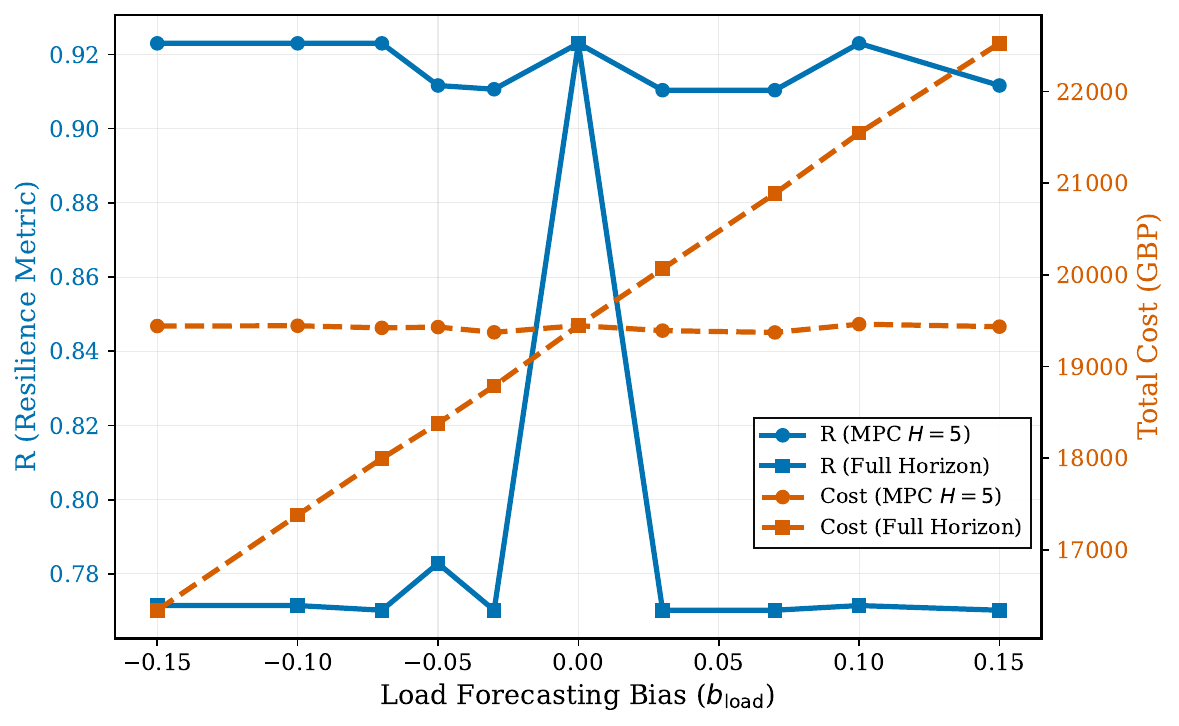}
    \caption{Effect of load forecasting bias}
    \label{fig:load_bias}
    \end{subfigure}
    \hfill
    \begin{subfigure}[b]{0.49\textwidth}
    \centering 
    \includegraphics[width=\textwidth]{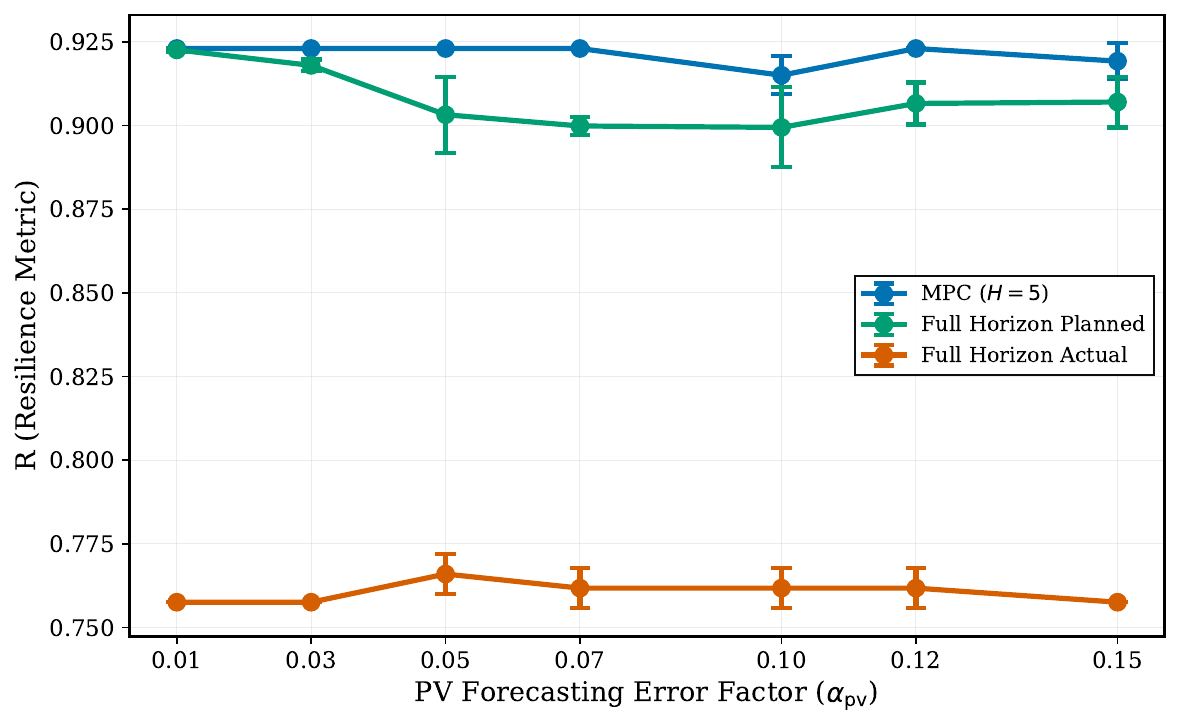}
    \caption{Effect of PV forecasting error}
    \label{fig:pv_error}
    \end{subfigure}
    \hfill
    \begin{subfigure}[b]{0.49\textwidth}
    \centering 
    \includegraphics[width=\textwidth]{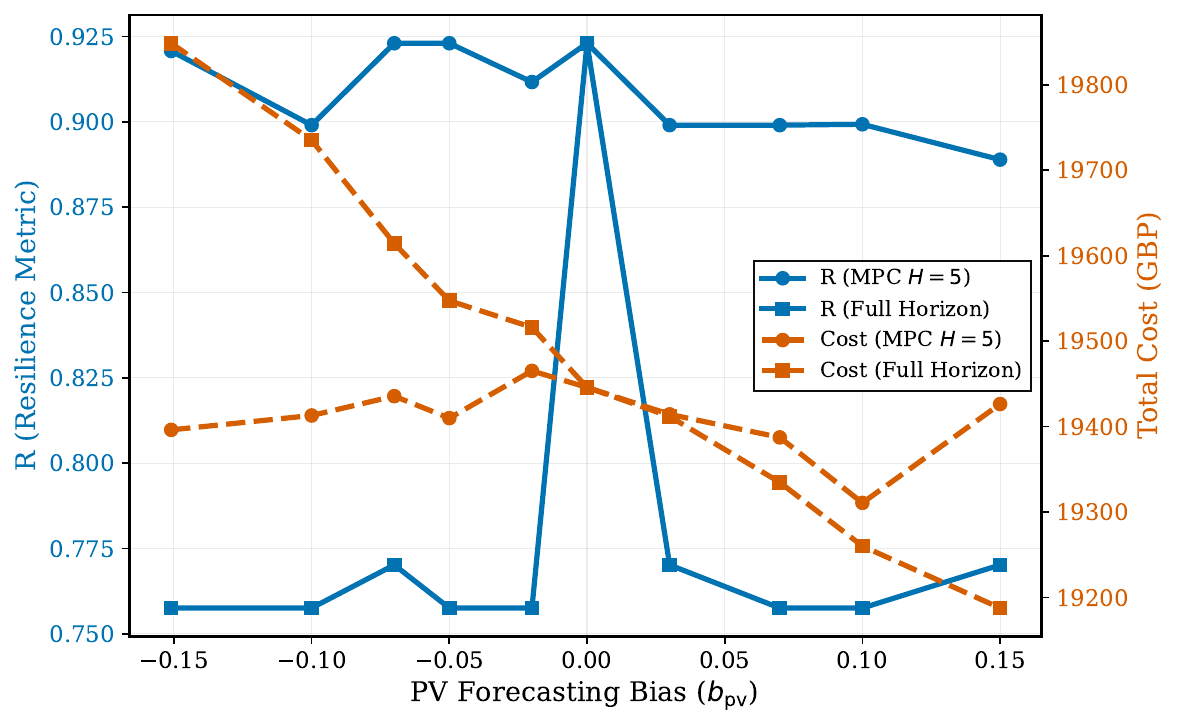}
    \caption{Effect of PV forecasting bias}
    \label{fig:pv_bias}
    \end{subfigure}
    \caption{MPC vs Full-Horizon under forecasting errors and systematic bias.}
    \label{fig:2}
\end{figure}

We now evaluate VPP robustness under realistic forecasting uncertainties. Figure \ref{fig:2} presents an analysis of how both random forecasting errors and systematic bias affect MPC and full-horizon optimization performance. The results in Figures \ref{fig:load_error} and \ref{fig:pv_error} represent averages over three independent runs with different random error realizations, with error bars indicating $\pm 1$ standard deviation to capture the stochastic nature of forecasting uncertainties.

\paragraph{Random Load Forecasting Errors}
Figure \ref{fig:load_error} shows system resilience under random load forecasting errors (with zero systematic bias, $b_{\text{load}}=0$) as a function of the error factor $\alpha_{\text{load}}$. The results reveal three distinct performance trajectories: MPC maintains consistently high resilience even as forecasting errors increase. The full-horizon planned solution, representing the theoretical performance if the pre-computed plan could be executed exactly, shows moderate degradation as forecasting errors increase, reflecting the inherent vulnerability of long-horizon planning to accumulating forecast uncertainties.

However, the full-horizon actual performance, representing the realized system performance when the pre-computed plan encounters real-world conditions and requires corrective actions, suffers severe degradation. This substantial performance gap between planned and actual full-horizon operation occurs because the pre-computed plan becomes increasingly infeasible as it encounters actual loads that deviate from forecasts. When control actions violate feasibility constraints (as detected by the checks in Figure \ref{fig:flowchart_full}), the system must abandon the globally optimized strategy and rerun optimization for the current time step only. This myopic re-optimization loses the critical look-ahead capability needed for resilience preparation, resulting in significantly reduced resilience during the actual outage period. The superior performance of MPC stems directly from its receding-horizon approach, which continuously adapts its strategy as forecast accuracy improves for near-term decisions.

\paragraph{Systematic Load Forecasting Bias}
Figure \ref{fig:load_bias} shows the impact of consistent load forecasting bias $b_{\text{load}}$ while setting random error to zero ($\alpha_{\text{load}}=0$).

For negative bias, where forecasted loads consistently underestimate actual demand, full-horizon charges batteries less aggressively based on the underestimated forecasts, so the pre-computed plan results in lower costs but also lower resilience (insufficient stored energy during grid outages). 

For positive bias, where forecasted loads consistently exceed actual demand, full-horizon charges batteries more aggressively than necessary, increasing operational costs. Despite this additional energy storage, resilience degrades. This occurs because during islanded operation, the pre-computed discharge plan assumes higher loads than actually occur. When planned discharge exceeds actual load, the excess power cannot be exported to the disconnected grid, causing feasibility violations. The corrective re-optimization makes myopic decisions that exhaust battery reserves early, leaving insufficient energy for later outage hours. Different corrective mechanisms could be considered to improve resilience in this scenario, though operational costs would remain elevated due to overcharging, and we leave such investigation to future work.

For both positive and negative bias, MPC maintains higher resilience because its continuous re-optimization with updated information allows it to adjust charging/discharging strategies as actual demand reveals itself.

\paragraph{PV Forecasting Errors and Bias}
Figures \ref{fig:pv_error} and \ref{fig:pv_bias} demonstrate that PV forecasting uncertainties produce similar patterns to load forecasting errors, with some directional differences. For random errors (Figure \ref{fig:pv_error}), MPC maintains robust resilience while full-horizon actual performance degrades as forecast uncertainty increases, mirroring the load error case. 

For systematic bias (Figure \ref{fig:pv_bias}), positive PV bias acts analogously to negative load bias, where overestimating available PV leads to insufficient battery pre-charging and lower full-horizon resilience. Negative PV bias mirrors positive load bias, where excess actual PV combined with planned battery discharge creates surplus power that cannot be exported during grid outages, causing feasibility violations and early battery depletion.
Across all PV forecasting scenarios, MPC maintains superior resilience through its adaptive re-optimization that dynamically adjusts both charging strategies and PV-battery coordination as actual PV generation reveals itself.

\subsection{Impact of Time-Varying Battery Efficiency}
\label{sec:sim_eff}

\begin{figure}[t!]
    \centering
    \includegraphics[width=0.5\textwidth]{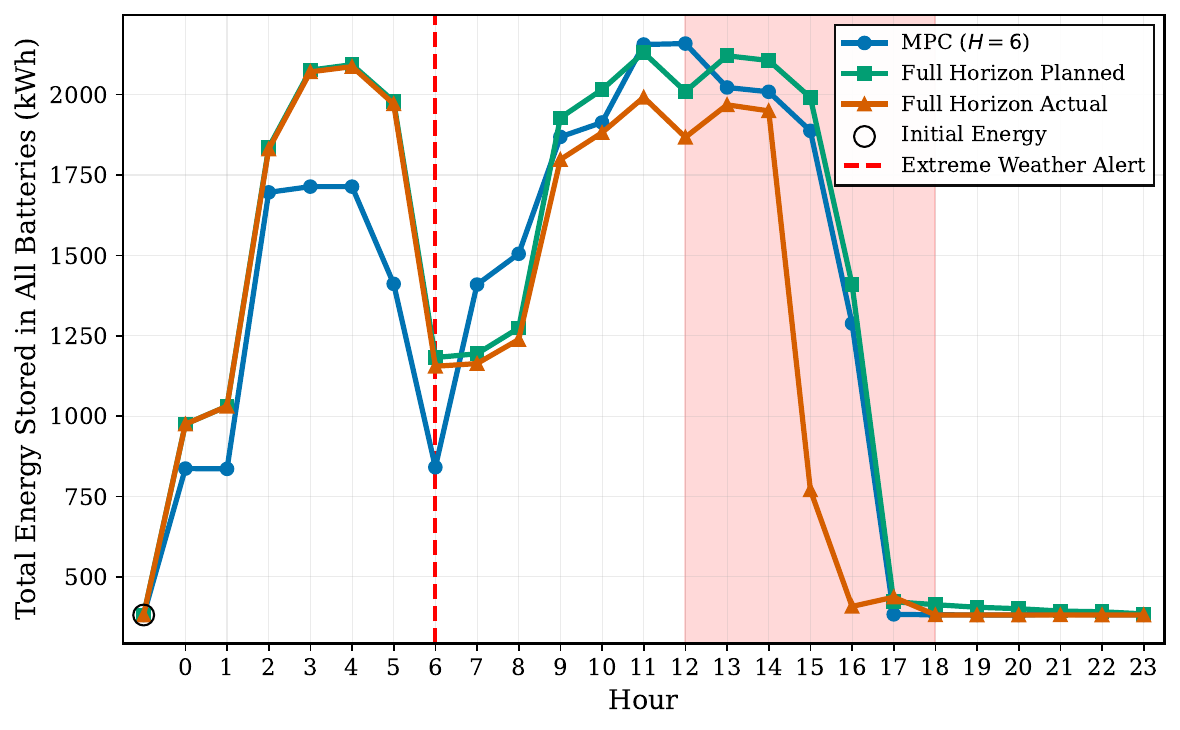}
    \label{fig:battery_eff}
    \caption{Total battery energy under time-varying battery efficiency.}
    \label{fig:3}
\end{figure}

\begin{table}[t!]
\centering
\caption{MPC vs Full-Horizon under time-varying battery efficiency.}
\label{tab:eff}
\begin{tabular}{lcc}
\toprule
 & \textbf{Resilience ($R$)} & \textbf{Total imported energy} \\
\midrule
\textbf{MPC} & 0.90 & 60854.07\,kWh \\
\textbf{Full Horizon (Actual)} & 0.81 & 60691.72\,kWh \\
\bottomrule
\end{tabular}
\end{table}

Next we evaluate system performance under time-varying battery efficiency, which represents a modeling uncertainty distinct from forecasting errors. For this experiment, we used a prediction horizon of $H=6$ hours for the MPC controller, enabling earlier extreme weather detection and providing additional time to pre-charge batteries and compensate for temperature-induced efficiency losses.

Figure \ref{fig:3} illustrates the total battery energy trajectories under time-varying efficiency conditions. The full-horizon optimization computes its complete 24-hour strategy using the initial nominal efficiency values. However, as this plan executes, charging operations generate internal heat, causing cell temperatures to rise and triggering efficiency degradation. The reduced actual charging efficiency means less energy is stored than planned. This compounding effect accumulates over the pre-charging period: at hour 11, immediately before the extreme weather event begins, the actual full-horizon stored energy falls approximately 150\,kWh below the planned level. During the extreme weather period, this energy shortfall reduces the ability to serve loads. Constrained by its pre-computed plan, the full-horizon approach cannot make adaptive adjustments to compensate for efficiency degradation and consequently under-charges its batteries, achieving $R=0.81$ as shown in Table \ref{tab:eff}.

In contrast, MPC maintains higher battery energy levels during the critical pre-charging and outage periods. Because MPC re-optimizes at each time step with updated battery efficiencies, when efficiencies degrade it adjusts subsequent charging strategies, increasing charging power to compensate for lower efficiency. This is reflected in Table \ref{tab:eff}, where MPC imports more energy from the grid. While this increases pre-event costs, it proves essential for achieving adequate backup capacity. This continuous adaptation enables MPC to achieve $R=0.90$, with $11\%$ higher resilience than full-horizon.

\section{Conclusions and Future Research}
\label{sec:conclusions}


This paper develops a framework for priority-aware load service and network-constrained scheduling in a technical VPP. The framework is designed to serve the most critical customers seamlessly while minimizing disruption to network operations during extreme weather, with a UK heatwave used as a case study. The framework is demonstrated over the entire extreme-weather timeline and fills a research gap by coupling a TVPP with a rolling-horizon controller that continuously operates before, during, and after an event, explicitly balancing resilience, cost, and renewable utilization under network constraints. 

The continuous receding-horizon control method, employed throughout the entire event timeline, provides superior performance under realistic operational conditions compared to traditional static planning approaches. MPC's fundamental advantage lies in its robustness to uncertainties. Since only the current time step decisions are implemented under actual conditions, optimization solutions are guaranteed to be feasible. Forecast errors affecting future decisions do not compromise current actions, as those decisions will be re-optimized with updated information in subsequent iterations. As conditions evolve and forecast accuracy improves for near-term events, the controller continuously refines its strategy. 

Quantitative results from simulations on the IEEE 33-bus network using July 2022 UK heatwave data validate these advantages. Under ideal conditions, MPC replicates the globally optimal solution ($R \approx 0.92$). Under realistic forecasting errors, systematic biases, and time-varying battery efficiency (conditions where VPPs must operate), MPC maintains robust performance at $R \approx 0.90\text{--}0.92$, while full-horizon optimization degrades to $R \approx 0.75\text{--}0.81$. This $11\text{--}20\%$ resilience improvement represents substantial practical value.

The ability to accommodate uncertainties through receding-horizon re-optimization demonstrates a key practical advantage: by continuously updating based on observed system behavior, MPC effectively handles model mismatches and multiple uncertainty sources. This makes MPC particularly valuable for real-world VPP implementations where perfect forecasts and models are unavailable, validating it as the preferred approach for extreme weather coordination.

\paragraph{Limitations of the approach}

The proposed framework functions under the assumption that the extreme weather would not result in a cascading or an eventual black-out of the network. The work assumes access to control the user-owned assets. The evaluation considers a single radial feeder without topology reconfiguration and the outage window and price signal are deterministic. In this work, the network constraints use a linearized power-flow model with AC power-flow checking, and forecast errors are modeled using parametric bias and horizon-dependent variance. Limitations include the lack of explicit modeling of communication delays and actuation limits.

\paragraph{Recommendations for Future Research}
Based on the limitations identified, future work can pursue three separate streams: (i) \textit{Uncertainty-aware MPC:} introduce stochastic or robust MPC with probabilistic outage timing and price forecasts, targeting resilience and voltage-violation risk. (ii) \textit{Multi-DER coordination:} integrate electric vehicles and wind generation, model availability and variability, and co-optimize real/reactive support under discrete inverter nameplate ratings. (iii) \textit{End-to-end deployment factors:} integrating real forecasting algorithms for load demand, renewable generation, and weather conditions into the MPC framework would enable more realistic assessment of system performance under actual forecasting uncertainties. Coupling state-of-the-art machine learning forecasters with the MPC controller, and accounting for communication and actuation delays, would provide more accurate characterization of forecast error impacts and enable co-optimization of forecasting and control strategies.

\section*{Acknowledgements}
This work was supported by the Engineering and Physical Sciences Research Council [grant number EP/Y005376/1] – VPP-WARD Project (\url{https://www.vppward.com}).


\bibliographystyle{unsrtnat}
\bibliography{references}

\end{document}